\documentclass[a4paper,11pt]{article}
\pdfoutput=1 

\usepackage{jcappub} 
\usepackage{graphicx}%
\usepackage{multirow}%
\usepackage{amsmath,amssymb,amsfonts}%
\usepackage{amsthm}%
\usepackage{mathrsfs}%
\usepackage{mathtools, slashed}
\usepackage[title]{appendix}%
\usepackage{xcolor}%
\usepackage{textcomp}%
\usepackage{manyfoot}%
\usepackage{booktabs} 
\usepackage{booktabs}%
\usepackage[numbers]{natbib}   
 \usepackage{tikz}
\usetikzlibrary{shapes.geometric, arrows.meta, positioning, calc}
\usepackage{algorithm}%
\usepackage{algorithmicx}%
\usepackage{algpseudocode}%
\usepackage{listings}%
\usepackage{mathtools, mathptmx}
\usepackage{breqn}
\usepackage{multicol}
\usepackage{enumerate}
\usepackage{dirtytalk}
\usepackage{subcaption}
\usepackage{caption}
\usepackage{rotating} 
\usepackage{microtype}
\usepackage{geometry}
\usepackage[utf8]{inputenc}
\usepackage[numbers]{natbib}
\usepackage{tikz}
\usepackage{feynmp-auto}
\usepackage[T1]{fontenc} 

\title{\boldmath Spontaneous Symmetry Breaking as a Late-Time Trigger for Interacting Dark Energy}

\author[a]{Pradosh Keshav MV}
\author[b]{NS Kavya}
\author[a]{Kenath Arun}

\affiliation[a]{Department of Physics and Electronics, CHRIST (Deemed to be University), Bengaluru 560029, India.}
\affiliation[b]{Center for Mathematical Needs, Department of Mathematics, CHRIST (Deemed to be University), Bengaluru 560029, India.}

\emailAdd{pradosh.keshav@res.christuniversity.in}
\emailAdd{kavya.ns@christuniversity.in}
\emailAdd{kenath.arun@christuniversity.in}

\abstract{Persistent tensions in the Hubble constant ($H_0$) and the matter clustering parameter ($S_8$) motivate late-time new physics that suppresses structure growth without significantly altering the background expansion history of the $\Lambda$CDM model. We study a new class of dark-sector dynamics in which a scalar dark energy (DE) field, governed by a $\mathbb{Z}_2$-symmetric quartic potential, interacts with dark matter (DM) through Yukawa and portal couplings. When the matter density drops below a critical threshold, a cosmological spontaneous symmetry breaking mechanism (SSB) generates a time-dependent vacuum expectation value $v(a)$ and activates an effective coupling $\beta(a)$. This produces a symmetric phase ($a \le a_c$) identical to $\Lambda$CDM at early times, and a broken phase ($a > a_c$) in which $\beta(a) > 0$ transfers energy from DM to DE, suppressing the linear growth of structure. We confront this framework with RSD, BAO, cosmic chronometers, and Pantheon+SH0ES supernova data, jointly with compressed \textit{Planck} distance priors, comparing a fixed $\Lambda$CDM background to a self-consistent coupled-scalar evolution. The RSD-only analysis shows a pronounced shift: the dynamical background yields $\Omega_m \approx 0.31 \pm 0.10$ and $\sigma_{8,0} \approx 0.59 \pm 0.01$, with a higher matter density but a substantially lower growth amplitude compared to the fixed-background case $\Omega_m \approx 0.20 \pm 0.09$ and $\sigma_{8,0} \approx 0.75 \pm 0.05$. In the full joint fit, we obtain $\Omega_m = 0.29 \pm 0.01$, $H_0 = 69.7 \pm 0.6~\mathrm{km\,s^{-1}\,Mpc^{-1}}$, and $\sigma_{8,0} = 0.78 \pm 0.01$. These results show that a late-time, SSB-activated interaction can efficiently damp structure growth and alleviate the $S_8$ tension while leaving the background expansion—and the inferred value of $H_0$ essentially unchanged. This mechanism provides a microphysical basis for the decoupling of structure growth from background expansion, suggesting distinct physical origins for the two tensions.}
\keywords{Interacting dark energy; spontaneous symmetry breaking; cosmological tensions; scalar–tensor theories; epoch-dependent coupling; cosmic structure growth }

\begin{document}
\maketitle
\flushbottom

\section{Introduction}

The microphysical origin and possible interactions of dark energy (DE) and dark matter (DM) remain key open questions in modern cosmology. Observations of Type Ia supernovae \cite{riess1998observational, perlmutter1999measurements}, the cosmic microwave background \cite{spergel2003firstyear, spergel2007threeyear}, and large-scale structure surveys \cite{tegmark2004cosmological, abazajian2004first, abazajian2005third} firmly establish that these two dark components dominate the cosmic energy budget. Within the concordance $\Lambda$CDM framework, where DE is modelled as a cosmological constant and DM as a cold, collisionless fluid, the expansion history and structure formation are well reproduced but remain phenomenological in nature. This lack of microphysical description not only precludes insight into potential dark sector dynamics but also becomes increasingly pressing in light of growing observational tensions. These include the $\sim5\sigma$ Hubble constant discrepancy \cite{riess2022comprehensive,Kavya:2025vsj,Mishra:2025rhi,Sudharani:2025cii}, the $2$–$3\sigma$ suppression of the matter clustering amplitude $S_8$ \cite{heymans2021kids}, and the cosmic coincidence problem \cite{mangano2003coupled}. Such tensions motivate extensions to $\Lambda$CDM that can couple background expansion and structure growth in a physically motivated way. Interacting dark energy (IDE) models \cite{di2017can, di2020interactingde, cai2005cosmic, olivares2006interacting, barrow2006interactions} provide one such framework, allowing energy and momentum exchange between the dark components through a microphysical kernel derived from a field-level Lagrangian.

While an interacting dark sector can, in principle, modify the late-time expansion history and suppress structure growth, potentially alleviating the $H_0$ and $S_8$ tensions \cite{quartin2008, bean2008}—most realizations remain purely phenomenological. Typically, the interaction kernel $Q$ is postulated as an \textit{ad hoc} function of background densities, without a clear microphysical origin. This arbitrariness has tangible consequences; certain forms of $Q$ trigger fatal instabilities in cosmological perturbations \cite{valiviita2008large}, while others are now significantly constrained by the combined CMB and large-scale structure data \cite{forconi2024double}. A deeper limitation, however, lies in the rigid link that these models impose between the background and perturbation evolution through the same $Q$ function. Any attempt to adjust the background energy flow inevitably modifies the linear growth source term, leading to correlated shifts in $H_0$ and $\sigma_8$. This degeneracy can be broken in frameworks where the coupling is globally suppressed in the early universe and becomes dynamically activated only at late times \cite{lin2023dark, garcia2024interacting,Sudharani:2024fdm}. In such scenarios, the background expansion remains nearly indistinguishable from the $\Lambda$CDM, while the coupling produces dominant effects on structure growth. This provides a natural theoretical basis for a key idea, emerging from recent analyses \cite{clark2023h, lucca2021dark, naidoo2024dark, benisty2024late, yashiki2025toward,Mishra:2025kzu}, that the $H_0$ and $\sigma_8$ tensions may originate from distinct physical mechanisms acting at different cosmic epochs. Late-time interactions like these naturally realize that separation in practice, and it is this decoupled regime that motivates the present work.

A natural realization of such a late-time, dynamically activated coupling arises in models where dark energy emerges from a spontaneous symmetry breaking (SSB), inducing a global phase transition in the dark sector. In these scenarios, the symmetry breaking is triggered not by local environment but by cosmic time or the background density itself, providing a cosmological—rather than astrophysical—mechanism for modifying the expansion or growth history. For instance, Banihashemi, Khosravi, and Shirazi~\cite{PhysRevD.101.123521} proposed a scenario in which the cosmological constant undergoes an Ising-like transition between two vacuum states at a critical redshift, while Garny \emph{et al.}~\cite{garny2024hot} introduced the Hot New Early Dark Energy (Hot-NEDE) model, where an SU($N$) dark-sector symmetry is supercooled and subsequently broken, releasing dark radiation prior to recombination. Vacuum-phase transition scenarios~\cite{di2018vacuum} and Ginzburg–Landau formulations of dark-energy critical phenomena~\cite{banihashemi2019ginzburg} likewise posit globally timed transitions in the vacuum energy density, whereas mirror-symmetry–breaking models~\cite{tan2019dark} interpret DE as a remnant of spontaneous mirror-sector SSB at the electroweak scale. Although these frameworks share the broad theme of global or epochal symmetry breaking, they generally prescribe fixed transition epochs or temperature thresholds without directly coupling the symmetry breaking to the evolving cosmic matter density. This leaves open the possibility of a dynamically triggered phase transition—analogous in spirit to environmentally dependent scalar-field mechanisms—that could reproduce the late-time phenomenology of global SSB models while preserving a nearly $ \ Lambda$CDM background expansion.

We now turn to the cosmological realization of this idea. Building on our earlier work \cite{V:2025oex}, we ask whether late-time growth observables can retain detectable signatures of such dynamically activated couplings. The mechanism parallels density-dependent screening models such as the chameleon \cite{khoury2004chameleon} and symmetron \cite{hinterbichler2010screening, hinterbichler2011symmetron,Sudharani:2024qnn}, where interactions are suppressed in dense environments and activated once the scalar acquires a nonzero vacuum expectation value (VEV). Similar density-triggered effects have appeared in dark-sector and modified-gravity studies \cite{pietroni2005dark, olive2008environmental, baldi2011clarifying, baldi2010hydrodynamical}, suggesting that perturbations can carry stronger observational signatures than the background expansion. However, for a globally timed transition in the vacuum energy density, this suppression implies that the background DM density dynamically triggers SSB. As the DM density redshifts below a critical value, the effective scalar mass changes sign, initiating SSB and turning on the coupling only in the late universe. The resulting effective coupling $\beta(a)$ is controlled by the scalar order parameter—vanishing in the symmetric phase and rising smoothly afterward. Interestingly, parametrizations of this type can be viewed as the low-energy limit of a density-induced SSB in a quartic scalar potential \cite{Morozumi2011Quantum}, which directly links the time dependence of the coupling to fundamental Lagrangian parameters (see also \cite{gubitosi2013effective, bloomfield2013dark, gleyzes2014healthy, gleyzes2015effective, gleyzes2016effective, koivisto2005growth}). This provides a natural explanation for why dark-sector interactions could have remained hidden until recently, while leaving potentially observable imprints in the growth of cosmic structure \cite{koivisto2008dynamics}.

In this work, we present a microphysical realization of this scenario using a scalar DE field with a quartic $\mathbb{Z}_2$-symmetric potential. This field is coupled to DM either through a Yukawa interaction (for fermionic DM) or via a scalar-portal term (for bosonic DM).  We assume that this coupling is dynamically activated only at late times, thereby alleviating the typical decoupling and fine-tuning issues that often affect interacting dark-sector models, particularly those involving large mass hierarchies between their components \cite{DAmico2016, Farrar2004, d2016quantum}. As the background matter density redshifts below a critical threshold, the scalar undergoes SSB, acquiring a time-dependent VEV $v(a)$ that induces an effective, epoch-dependent coupling $\beta(a)$ rising smoothly at late times. Consequently, the interaction strength emerges as a symmetry-protected, radiatively stable feature of the underlying dynamics, in contrast to the \textit{ad hoc} or environment-dependent descriptions typical of earlier phenomenological models~\cite{sin1994late, ji1994late, brax2009decoupling, bamba2013spontaneous, zhang2020obtaining}. At the cosmological level, the field equations reduce to an effective energy–momentum exchange kernel $Q(a)$ that vanishes during the symmetric phase. This kernel activates only after symmetry breaking, causing the background expansion to closely resemble that of the \(\Lambda\)CDM model. To confront our model with observations, we combine \textit{Planck}-compressed distance priors—which anchor the early-universe geometry—with late-time probes, including BAO, cosmic chronometers, Pantheon+SH0ES supernovae, and redshift-space distortions. Together, these datasets test whether an SSB-induced coupling can produce dataset-dependent shifts in growth parameters while leaving the background expansion $H(z)$ effectively unchanged.

This paper is organized as follows. In Section~2, we outline the theoretical framework of IDE models, reviewing the scalar–tensor origin of DM couplings and setting the background for the proposed model. Section~3 develops the linear perturbation theory for an epoch-dependent interaction and derives its implications for the growth of structure. Section~4 introduces the central mechanism—a SSB mechanism dynamically triggered by the cosmic matter density—which provides the physical origin of the late-time coupling. In Section~5, we construct a compact phenomenological parametrization of the effective coupling $\beta(a)$ that reproduces the microphysical behaviour of the IDE model. Section~6 presents the observational analysis, confronting the model with a combination of low- and intermediate-redshift datasets to quantify the coupling strength and its effect on the $S_8$ tension. Finally, Section~7 discusses the broader implications and prospects, while technical derivations are provided in the Appendices.

Throughout the paper, we adopt the following convention, which we mention here for clarity. With our sign conventions \(Q>0\) implies energy transfer from DM to DE; background evolution is written in terms of cosmic time $t$, with overdots denoting  \(d/dt\) and $H=\dot{a}/a$. Perturbations are expressed in conformal time \(\tau\), with primes denoting \(d/d\tau\) and $\mathcal{H}=a'/a$ . This means that background quantities follow the usual cosmic-time Friedmann equations, while perturbations are treated consistently in the conformal Newtonian gauge.

\section{Scalar Field Dynamics and Interacting Dark Sector}
\label{sec:2}
\subsection{Scalar-Tensor origin of constant couplings}

We start with an action in scalar-tensor theories \cite{baldi2012multiple, amendola2004linear, amendola2008quintessence}, where a dynamic scalar field is nonminimally coupled to matter via a conformal rescaling of the metric in the Einstein frame:
\begin{equation}
S = \int d^4x \sqrt{-g} \left[ \frac{M_{\rm Pl}^2}{2} R 
- \frac{1}{2} g^{\mu\nu} \partial_\mu \phi \partial_\nu \phi 
- V(\phi) + \mathcal{L}_m\left( A^2(\phi) g_{\mu\nu}, \psi_m \right) \right],
\label{eq:action_scalar_tensor}
\end{equation}
where \( M_{\rm Pl} \) is the reduced Planck mass, \( \phi \) is a canonical scalar field with potential \( V(\phi) \), and \( \mathcal{L}_m \) denotes the Lagrangian of the matter fields \( \psi_m \). The function \( A(\phi) \) characterizes the conformal coupling between the scalar field and matter. The physical (Jordan-frame) metric experienced by matter is:
\begin{equation*}
\tilde{g}_{\mu\nu} = A^2(\phi) g_{\mu\nu}, \label{eq:conformal_metric}
\end{equation*}which induces a $\phi$-dependence of its physical mass and energy--momentum tensor \cite{bekenstein1993gravitational, wetterich2003probing}.

Variation of the action with respect to \( g_{\mu\nu} \) yields the Einstein equations in the usual form, while the matter fields couple directly to the rescaled metric \( \tilde{g}_{\mu\nu} \). Consequently, the matter energy--momentum tensor is covariantly conserved with respect to \( \tilde{g}_{\mu\nu} \), but not with respect to the Einstein-frame metric \( g_{\mu\nu} \). One finds the modified conservation law
\begin{equation}
\nabla^\mu T^{(m)}_{\mu\nu} =- \frac{d \ln A(\phi)}{d\phi} \, T^{(m)} \nabla_\nu \phi,
\label{eq:nonconservation_tensor}
\end{equation}
where \( T^{(m)} \equiv T^{\mu (m)}_{\ \mu} \) is the trace. Splitting the total energy–momentum tensor  \(\nabla^\mu T^{\text{(tot)}}_{\mu\nu} = 0\) into DM and DE components, we write
\begin{equation}
    \nabla^\mu T^{\rm (DM)}_{\mu\nu} = -Q_\nu, 
    \qquad
    \nabla^\mu T^{\rm (DE)}_{\mu\nu} = +Q_\nu,
    \label{eq:split_conservation}
\end{equation}
and adopt the commonly used choice \(Q_\nu = Q\,u_\nu\) (alignment with the DM four-velocity) to avoid spurious momentum transfer in the DM rest frame \cite{clemson2012interacting, pettorino2013testing}. 

For non-relativistic CDM \( T^{(m)} = -\rho_m \), the interaction vector is therefore:
\begin{equation}
Q_\nu =  -\nabla^\mu T^{\rm (DM)}_{\mu\nu} = +\frac{d \ln A(\phi)}{d\phi} \, \rho_m \, \nabla_\nu \phi,
\label{eq:interaction_vector}
\end{equation} and at the background level (\(\phi=\phi(t)\)) the temporal component reduces to
\begin{equation}
Q=\frac{\beta(\phi)}{M_{\rm Pl}}\dot\phi\rho_m,\qquad
\beta(\phi)\equiv M_{\rm Pl}\frac{d\ln A(\phi)}{d\phi}.
\label{eq:q_beta_form}
\end{equation}
Our sign convention is such that \(Q > 0\) indicates energy transfer from DM to DE, leading to faster DM dilution and increased sourcing of DE when \( \beta(\phi)\dot{\phi} > 0\). Substituting Eq.~\eqref{eq:q_beta_form} into the scalar field equation of motion yields the general Klein--Gordon equation:
\begin{equation}
    \ddot{\phi} + 3H \dot{\phi} + V'(\phi) = +\frac{\beta(\phi)}{M_{\rm Pl}} \rho_m, \label{eq:klieneqn}
\end{equation}consistent with the above sign choices.

The exponential conformal factor \( A(\phi) = \exp(\beta(\phi) / M_{\rm Pl}) \) results in a constant coupling \( \beta(\phi) = \beta \), which is the standard choice in coupled quintessence \cite{amendola2000coupled, amendola2014multifield}. More generally, field-dependent forms are naturally realized in effective field theory and string-inspired constructions \cite{koivisto2008dynamics, van2015disformal, damour1994string}, which are not within the scope of the current study.

\subsection{Background dynamics of IDE models}
In this section, we review the essential background dynamics of the interacting dark sector, which provide the foundation for the perturbative and phenomenological analyses developed in the following sections. Dark energy is modeled as a canonical scalar field $\phi(\mathbf{x},t)$ minimally coupled to gravity \cite{amendola2000perturbations, damour1990dark}, while dark matter interacts with this field through a conformal coupling encoded in a field-dependent mass. Such an interaction implies that the DM energy--momentum tensor is not separately conserved, leading to an exchange of energy and momentum between the two components at the background level.

Decomposing the scalar field into a homogeneous background and a small perturbation,
\begin{equation*}
    \phi(\mathbf{x},t)=\bar\phi(t)+\delta\phi(\mathbf{x},t),
\end{equation*}
and using the interaction kernel derived in Eq.~\eqref{eq:q_beta_form}, the background continuity equations become
\begin{align}
    \dot{\bar\rho}_{\rm DM}+3H\bar\rho_{\rm DM} &= -Q, \\
    \dot{\bar\rho}_{\phi}+3H(1+w_\phi)\bar\rho_{\phi} &= +Q,
    \label{eq:de_bg_corrected}
\end{align}
with the compact source term
\begin{equation}
    Q=\frac{\beta(a)}{M_{\rm Pl}}\,\dot{\bar\phi}\,\bar\rho_{\rm DM}.
    \label{eq:Q_background}
\end{equation}
Hence, for $Q>0$, DM dilutes faster than the standard $a^{-3}$ scaling while DE is sourced, and $Q<0$ slows the dilution of background DM energy density $\bar\rho_{\rm DM}$.

The dark energy background density and pressure are given by the standard minimally coupled relations,
\begin{equation*}
    \bar\rho_{\phi}=\tfrac{1}{2}\dot{\bar\phi}^2+V(\bar\phi),\qquad
    \bar p_{\phi}=\tfrac{1}{2}\dot{\bar\phi}^2-V(\bar\phi),
\end{equation*} so that the scalar field contributes to the total energy budget in the same form as a canonical quintessence component. At this level, the coupling to DM enters only through the energy-exchange term $Q$, which is proportional to $\dot{\bar\phi}$ and hence strongly suppressed when the field evolves slowly. However, once perturbations are introduced, the interaction term becomes spatially inhomogeneous, producing corrections to the continuity and Euler equations of both the DM and scalar sectors. Substituting \eqref{eq:Q_background} into total energy conservation yields the background Klein--Gordon equation,
\begin{equation}
    \ddot{\bar\phi}+3H\dot{\bar\phi}+V'(\bar\phi)=\frac{\beta(a)}{M_{\rm Pl}}\,\bar\rho_{\rm DM}.
    \label{eq:KG_background_corrected}
\end{equation}
\medskip

In this work, we treat the coupling as an epoch-dependent function $\beta=\beta(a)$ (equivalently $\beta(a)=\beta[\bar\phi(a)]$ along the background trajectory). This parametrization allows the coupling to remain negligible at early times and activate gradually at later epochs. Compared with purely phenomenological $Q(a)$ prescriptions, promoting the coupling to a time-dependent, conformally motivated form retains a clear connection to the underlying scalar--tensor Lagrangian while permitting controlled departures from constant-$\beta$ models.

Together with the Friedmann equations,
\begin{equation*}
    3M_{\rm Pl}^2H^2=\bar\rho_{\rm tot},\qquad
    \dot H=-\frac{1}{2M_{\rm Pl}^2}(\bar\rho_{\rm tot}+\bar p_{\rm tot}),
\end{equation*}
Eqs.~\eqref{eq:de_bg_corrected}--\eqref{eq:KG_background_corrected} recover the standard coupled-quintessence background in the constant-$\beta$ limit while allowing epoch-dependent activation of the interaction, which becomes central to the growth analysis in the subsequent sections.

\section{Linear Perturbations with Epoch-Dependent Coupling}
\label{sec:linear_perturbations}

Linear perturbations of the interacting dark sector are formulated in the conformal Newtonian gauge \cite{amendola2004linear}, where the time-dependent coupling drives deviations from the standard \(\Lambda\)CDM evolution of density and velocity fields. The background geometry is assumed to be spatially flat FLRW:
\begin{equation}
    ds^2 = a^2(\tau)\left[-(1 + 2\Phi)\,d\tau^2 + (1 - 2\Psi)\,\delta_{ij}\,dx^i dx^j\right], \label{eq:metricnewton}
\end{equation}
where \( a(\tau) \) is the scale factor and \( \Phi,\Psi \) represents scalar potentials. At the background level, we set \( \Phi = \Psi = 0 \) and neglect anisotropic stress \cite{ma1995cosmological}. We describe perturbations in terms of the density contrast and velocity divergence,
\begin{equation*}
    \delta_i \equiv \frac{\delta\rho_i}{\bar{\rho}_i}, 
    \qquad 
    \theta_i \equiv \nabla \cdot \vec{v}_i,
    \label{eq:delta_theta_def}
\end{equation*}
where overbars denote background quantities and \(\vec{v}_i\) is the peculiar velocity of component \(i\). Primes denote derivatives with respect to conformal time, \('\equiv d/d\tau'\), and the Fourier-space convention \(\theta = i\mathbf{k}\!\cdot\!\mathbf{v}\) is adopted.

We define the energy-transfer rate \(Q\) in cosmic time \(t\) as in Eq.~\eqref{eq:q_beta_form},
\begin{equation}
     Q(t,\mathbf{x}) \equiv \frac{\beta(\phi)}{M_{\rm Pl}}\,\rho_{\rm DM}(t,\mathbf{x})\,\dot\phi(t,\mathbf{x}),
\end{equation}where dots \(d/dt\). In the conformal-time perturbation equations (primes \(d/d\tau\)) the source enters multiplied by a factor \(a\); we therefore define the conformal-time source
\begin{equation}
    \mathcal{Q}(\tau,\mathbf{x}) \equiv a(\tau)\,Q(t(\tau),\mathbf{x}) = \frac{\beta(a)}{M_{\rm Pl}}\,\bar\rho_{\rm DM}(\tau)\,\phi'(\tau,\mathbf{x}),
    \label{eq:Q_conf_def}
\end{equation}
where \(\phi' = a\dot\phi\) and \(\mathcal{H}=a'/a\), all conformal-time perturbed equations below are written in terms of \(\mathcal Q\) and \(\delta\mathcal Q\). 

In the general IDE formulation, the coupling $\beta$ is a function of the scalar field, $\beta = \beta(\phi)$, so that perturbations in $\phi$ induce fluctuations in the coupling itself. For phenomenological analyses, however, it is often convenient to express the background dependence as $\beta(a) \equiv \beta[\bar{\phi}(a)]$, treating the time evolution through the background field. In this case, linearization around the homogeneous background yields a first-order correction $\delta\beta = \beta_{,\phi}\,\delta\phi$, which must be retained whenever $\beta$ depends explicitly on $\phi$. We explicitly indicate below the terms in which $\beta_{,\phi}$ contributes, distinguishing them from the purely time-dependent $\beta(a)$ case. The full linearized expression, including metric contributions to $\delta\mathcal{Q}$, is provided in Appendix~\ref{app:deltaQ}.

At first order, the canonical scalar-field fluctuation in Fourier space obeys the perturbed Klein–Gordon equation. With our conformal-time conventions, a convenient form (consistent with \cite{ma1995cosmological, sawicki2013consistent, amendola2004linear} ) is:
\begin{multline}
    \delta\phi'' + 2\mathcal{H}\,\delta\phi' + \big(k^2 + a^2 V''(\bar\phi)\big)\delta\phi
    - \bar\phi'\big(\Phi' + 3\Psi'\big) + 2a^2 V'(\bar\phi)\,\Phi \\
    = a^2\frac{\beta(\bar\phi)}{M_{\rm Pl}}\,\bar\rho_{\rm DM}\big(\delta_{\rm DM} + \Phi\big)
    \;+\; a^2\frac{\beta_{,\phi}(\bar\phi)}{M_{\rm Pl}}\,\bar\rho_{\rm DM}\,\delta\phi.
     \label{eq:KG_perturbed}
\end{multline}
A few remarks are in order: (i) the metric-sourcing combination on the left-hand side involves \(\Phi'\) and \(\Psi'\); if anisotropic stress is neglected, then \(\Psi=\Phi\) and the combination reduce accordingly, (ii) the first term on the right-hand side is the direct matter-sourced coupling (present even for \(\beta=\mathrm{const}\)), and (iii) the last term appears only when \(\beta=\beta(\phi)\) and captures the linear response of the coupling to field perturbations. 

Using the conformal-time source $\mathcal{Q}$, the linearized interaction perturbation is, to first order,
\begin{equation}
    \delta\mathcal{Q}
    \;=\;\frac{\beta(\bar\phi)}{M_{\rm Pl}}\,\bar\rho_{\rm DM}\big(\bar\phi'\,\delta_{\rm DM} + \delta\phi'\big)
    \;+\;\frac{\beta_{,\phi}(\bar\phi)}{M_{\rm Pl}}\,\bar\rho_{\rm DM}\,\bar\phi'\,\delta\phi
    \;+\; \text{(metric piece)}.
     \label{eq:deltaq_conf}
\end{equation}
The first two terms follow directly from linearizing $\mathcal{Q}=(\beta/M_{\rm Pl})\,\rho_{\rm DM}\,\phi'$ about the background, with the final term collecting standard metric contributions (see Appendix~\ref{app:deltaQ} for the full expression). When adopting the phenomenological ansatz $\beta=\beta(a)$, i.e.\ treating the coupling as a time-dependent function rather than an explicit field dependence, the $\beta_{,\phi}$ term is absent and $\beta(\bar\phi)$ is replaced by $\beta(a)$. In this formulation, the background expansion remains unaltered, while the time variation of $\beta(a)$ introduces modifications entirely at the perturbative level through $\delta\mathcal{Q}$ and the momentum-exchange terms via the evolution of inhomogeneities rather than changes in the homogeneous background \cite{bean2004probing, de2010measuring}.

Inserting the split conservation \(\nabla^\mu T^{\rm (DM)}_{\mu\nu}=-Q_\nu\) (with the customary DM-frame choice \(Q_\nu=Q u_\nu^{\rm (DM)}\)) into the perturbed continuity equation (in conformal time) yields:
\begin{equation}
    \delta_{\rm DM}' + \theta_{\rm DM} - 3\Psi' 
    = \frac{\mathcal{Q}}{\bar{\rho}_{\rm DM}}\,\delta_{\rm DM} + \frac{\delta\mathcal{Q}}{\bar{\rho}_{\rm DM}}.
    \label{eq:dm_continuity_expanded_conf}
\end{equation}The first term on the right-hand side hence reduces the growth of DM overdensities by removing background energy from the DM sector, while the second term represents the spatially inhomogeneous modulation of the coupling sourced by fluctuations. The corresponding Euler equation for DM (again in the DM-frame \(Q_\nu=Q u_\nu\)) is
\begin{equation}
  \theta_{\rm DM}' + \mathcal{H}\,\theta_{\rm DM} - k^2\,\Phi 
    = \frac{\mathcal{Q}}{\bar{\rho}_{\rm DM}}\,\theta_{\rm DM},
    \label{eq:dm_euler_expanded_conf}
\end{equation}
so that the interaction acts as an additional velocity-damping (or acceleration) term depending on the sign of \(\mathcal Q\). (If an alternative momentum-transfer prescription is chosen, e.g.~\(Q_\nu\propto u_\nu^{\rm (DE)}\), the right-hand side structure changes and may introduce relative-velocity terms.)

For the DE sector, treated phenomenologically as an imperfect fluid with equation of state \(w_{\rm DE}\) and effective sound speed \(c_s^2\), the linearized continuity and Euler equations read respectively as:
\begin{equation}
   \delta_{\rm DE}' + (1 + w_{\rm DE})(\theta_{\rm DE} - 3\Psi') 
    + 3\mathcal{H}(c_s^2 - w_{\rm DE})\,\delta_{\rm DE} 
    = \frac{\mathcal{Q}}{\bar{\rho}_{\rm DE}}\,\delta_{\rm DE} + \frac{\delta\mathcal{Q}}{\bar{\rho}_{\rm DE}},
    \label{eq:de_continuity_expanded_conf}
\end{equation}
\begin{equation}
\theta_{\rm DE}' + \mathcal{H}(1 - 3w_{\rm DE})\,\theta_{\rm DE} 
    - \frac{c_s^2\,k^2}{1 + w_{\rm DE}}\,\delta_{\rm DE} 
    = \frac{\mathcal{Q}}{\bar{\rho}_{\rm DE}}\,\theta_{\rm DE}.
\label{eq:de_euler_corrected_conf}
\end{equation}Since Eqs \eqref{eq:de_continuity_expanded_conf}--\eqref{eq:de_euler_corrected_conf} are written at the fluid level, they allow for an independent specification of \(w_{\rm DE}\) and \(c_s^2\) as separate phenomenological quantities. As noted by \cite{pettorino2008coupled}, this freedom accommodates a wide range of clustering behaviors, though not every \((w_{\rm DE},c_s^2)\) combination corresponds to a consistent scalar-field realization. For instance, consider the case of a canonical scalar field \(\phi\), the first-order perturbations of its energy density, pressure, and momentum flux are given by
\begin{align*}
\delta\rho_\phi &= \frac{1}{a^2}\Big(\bar\phi'\,\delta\phi' - (\bar\phi')^2\,\Phi\Big) + a^2 V'(\bar\phi)\,\delta\phi, \\
\delta p_\phi   &= \frac{1}{a^2}\Big(\bar\phi'\,\delta\phi' - (\bar\phi')^2\,\Phi\Big) - a^2 V'(\bar\phi)\,\delta\phi,
\end{align*}
\begin{equation*}
(\bar\rho_\phi+\bar p_\phi)\,\theta_\phi \;=\; \frac{k^2}{a^2}\,\bar\phi'\,\delta\phi,
\end{equation*}
From these relations, one finds that the rest-frame sound speed of a canonical scalar is unity, \(c_{s,{\rm rest}}^2=1\), and that the field carries no intrinsic anisotropic stress. Consequently, the fluid equations (\ref{eq:de_continuity_expanded_conf})–(\ref{eq:de_euler_corrected_conf}) are precisely equivalent to the perturbed Klein–Gordon equation (\ref{eq:KG_perturbed}) {only when} the fluid parameters satisfy these scalar-field relations—namely, when \(c_s^2\to1\) in the rest frame and the anisotropic stress vanishes.

At this point, it is helpful to clarify the dynamical regime considered in this work. In the limit discussed above, the background interaction term 
\(Q \!\propto\! \beta(a)\,\dot{\bar\phi}\,\rho_{\mathrm{DM}}\) has a negligible impact on the homogeneous expansion history. This is because the background field remains nearly frozen, \(\dot{\bar\phi}\!\approx\!0\), prior to symmetry breaking in the dark sector and evolves only mildly thereafter.  As a result, the expansion rate is effectively indistinguishable from that of \(\Lambda\)CDM, with fractional corrections \(\Delta H/H \lesssim 10^{-3}\) for benchmark values \(\beta_0 \lesssim 0.1\) (see also Sec.~\ref{sec:adiabatic} and Sec.~\ref{sec:backgroundQ}).

The same coupling, however, enters the first-order perturbation equations through \(\delta\mathcal{Q}\) and the modified gravitational coupling, leading to nontrivial effects on structure formation. This regime can therefore be described as \emph{quasi-adiabatic}: the scalar field is dynamically inert at the background level but mediates an unsuppressed fifth force at the perturbative level. Unlike symmetron or chameleon models, where screening suppresses local interactions in high-density environments, here the suppression arises purely from the cosmological slow-roll of the background field, allowing for a frozen background yet an active coupling in the perturbations. This paper focuses on exploring the phenomenological consequences of this eccentric regime using the late-time benchmarks described in Sec.~\ref{sec:cospara} and tested against data in Sec.~\ref{sec:observationalconstraints}.

\section{Spontaneous Symmetry Breaking and Epoch-Dependent Coupling}
\label{sec:microphysics}
Spontaneous symmetry breaking (SSB) has long been recognized as a unifying mechanism driving cosmological phase transitions \cite{kirzhnits1972macroscopic, kolb1990origin}, emergent phenomena in condensed matter systems \cite{nambu2009nobel, anderson1963plasmons}, and mass generation in particle theory \cite{englert1964broken, weinberg1967model}. A common feature of these settings is that interactions are naturally suppressed in the symmetric phase and become operative only after the symmetry is broken. In the present context, this logic translates into a dynamical activation of the dark-sector coupling. When the universe is dense, the scalar sits in the symmetric phase, and the DE–DM coupling is negligible. But once the background density falls below a critical threshold, the scalar acquires a vacuum expectation value (VEV), and the interaction is dynamically activated. This transition is thus the cosmic analogue of the Higgs mechanism, with the expansion of the universe playing the role of thermal cooling. In this section, we show explicitly how a time-varying VEV induced by an SSB potential manifests in the DM sector and how that maps onto the \(\beta(a)\) parametrization used in the subsequent cosmological analysis.

\subsection{Field-Theoretic Origin of the Epoch-Dependent Coupling}

We begin with a minimal Einstein-frame action in Eq.~\eqref{eq:action_scalar_tensor}, with the DM sector singled out:
\begin{equation}
S = \int d^4x \sqrt{-g} \left[ \frac{M_{\rm Pl}^2}{2} R - \frac{1}{2} g^{\mu\nu} \partial_\mu \phi \, \partial_\nu \phi - V(\phi) \right] + S_{\rm DM}\!\left[\Psi, \tilde{g}_{\mu\nu}\right],
\label{eq:total_action}
\end{equation}
where $V(\phi)$ is the scalar potential and $\Psi$ denotes DM fields. Recall that in Eq \eqref{eq:conformal_metric}, if DM particle masses are $\phi$-dependent, $m(\phi)=A(\phi)m_0$, so that the cosmological coupling is the logarithmic mass slope:
\begin{equation}
\beta_{\rm DM}(\phi) \equiv M_{\rm Pl}\,\frac{m'(\phi)}{m(\phi)} = M_{\rm Pl}\,\frac{A'(\phi)}{A(\phi)}.
\label{eq:beta_dm}
\end{equation}
If $A(\phi)=\exp(c\,\phi/M_{\rm Pl})$, one obtains $\beta(\phi)=c$, recovering the coupled-quintessence limit.

Assuming that $\beta(\phi)$ becomes explicitly time-dependent once the DE scalar undergoes spontaneous symmetry breaking, the field relaxes to a background configuration $\phi(\tau)\simeq v(a)$ whose VEV evolves with the cosmic background. As a result, even a constant microscopic slope $d\ln A/d\phi$ induces a macroscopic scale-factor dependence, $\beta(a)\equiv\beta(\bar\phi(a))$. This mechanism is familiar from screening and SSB constructions \cite{hinterbichler2011symmetron, khoury2004chameleon, olive2008environmental}, where environment-dependent VEVs generate effective interactions that vary across epochs, and has been shown to leave observable imprints on growth and large-scale structure \cite{brax2011linear,damour1994string}. The underlying discrete $\mathbb{Z}_2$ symmetry (or, in UV completions, a softly broken $U(1)$) protects the potential and mediates the long-range.

The DE scalar potential is chosen as a Higgs‐type quartic:
\begin{equation}
V(\phi) = \frac{\lambda}{4} \left( \phi^2 - v^2 \right)^2,
\label{eq:higgs_potential}
\end{equation}
with $\lambda>0$ and a broken-phase VEV.  As the universe expands and the background density decreases, the effective mass term changes sign at a critical scale factor $a_c$, displacing the minimum away from $\phi=0$. This evolving order parameter directly feeds into the DM sector, since DM couplings render the physical mass an explicit function of $v(a)$.

Quartic symmetry-breaking potentials of this type, such as Higgs-portal and dark-singlet constructions, link quartic potentials to DM mass generation \cite{landim2018dark}, and classically scale-invariant models generally demonstrate how radiative symmetry breaking in quartic sectors can yield dynamical VEVs in hidden fields \cite{altmannshofer2015light}. However, once the scalar settles into its symmetry-breaking minimum, it is standard to expand around the background configuration, $\phi=v(a)+\delta\phi$. This expansion diagonalizes the fluctuations, defines the scalar mass, and makes explicit the residual mediator couplings that enter the DM Lagrangian. To illustrate this mechanism, we consider two simple and widely studied cases of DM couplings: 

\paragraph{Yukawa dark matter (fermion):}  
Consider a fermionic DM field with Lagrangian $\mathcal{L}_{\rm DM}=\bar{\psi}(i\slashed{\nabla}-m_\psi(\phi))\psi$ and $\phi$-dependent mass $m_\psi(\phi)=m_0+y\phi$ \cite{patt2006higgs}. Expanding around the symmetry-breaking minimum $\phi=v+\delta\phi$ gives:
\begin{equation}
m_\psi(\phi)=m_\psi(v)+y\,\delta\phi,\qquad m_\psi(v)=m_0+y\,v,
\label{eq:fermion_mass_shift}
\end{equation}
so that the physical fermion mass tracks the evolving VEV $v(a)$ while the mediator — DM vertex is $g_{\psi\psi\phi}=y$. The corresponding cosmological coupling is:
\begin{equation}
\beta(a)=\beta_{\rm DM}(\bar{\phi})\big|_{\bar{\phi}\simeq v(a)}
=\frac{y\,M_{\rm Pl}}{m_0+y\,v(a)} .
\label{eq:beta_yukawa_time}
\end{equation}
Numerically, taking $m_0\simeq100\,{\rm GeV}$ \cite{PhysRevLett.118.141802}, $y\simeq2\times10^{-17}$, and $v/M_{\rm Pl}\sim10^{-5}$ with the reduced Planck mass $M_{\rm Pl}=2.4\times10^{18}\,{\rm GeV}$, one finds $yM_{\rm Pl}\simeq 49\,{\rm GeV}$ and $y v\ll m_0$, so that Eq.~\eqref{eq:beta_yukawa_time} yields $\beta_0\approx0.5$ today. In the cosmological settings, however, these magnitudes enter the background exchange term $Q=\beta\,\bar\rho_{\rm DM}\,\dot\phi/M_{\rm Pl}$ and the perturbation source term $\mathcal{Q}=aQ$.

\paragraph{Scalar portal dark matter:} 
As an alternative to fermionic Yukawa couplings, DM may couple to the DE scalar through a scalar portal, \(\mathcal{L}_{\rm int}=g\,\phi^2\chi^2\) (or $\zeta\,\phi\,\chi^2$) \cite{arcadi2018dark}. 
In this case, the dark scalar mass depends quadratically on the DE field: 
\begin{equation}
m_\chi^2(\phi)=m_{0}^2+g\,\phi^2 \;\Rightarrow\; 
m_\chi^2(v+\delta\phi)=m_{0}^2+g\,v^2+2g\,v\,\delta\phi+g\,(\delta\phi)^2,
\label{eq:scalar_mass_shift}
\end{equation}
so that expansion about the SSB vacuum induces both a shifted physical mass and residual mediator couplings, $g_{\chi\chi\phi}=2g\,v$ and $g_{\chi\chi\phi\phi}=g$. 
The cosmological coupling, therefore, tracks the time dependence of the VEV: 
\begin{equation}
\beta(a)=M_{\rm Pl}\,\frac{d\ln m_\chi(\phi)}{d\phi}\Big|_{\phi=v(a)}=\frac{M_{\rm Pl}\,g\,v(a)}{m_{0}^2+g\,v(a)^2},
\label{eq:beta_portal_time}
\end{equation}
and makes it explicit that the effective interaction strength is modulated by the ratio of the evolving order parameter $v(a)$ to the physical DM mass scale. For $m_0^2\gg g v(a)^2$, the coupling is strongly suppressed, reproducing $\Lambda$CDM–like phenomenology, while for lighter portal DM, the growth of $v(a)$ at late times can source a non-negligible $\beta(a)$.

\medskip

Note that Eqs~\eqref{eq:beta_yukawa_time} and \eqref{eq:beta_portal_time} illustrate the central mechanism of the framework: in both Yukawa and scalar-portal realizations, the DM mass function $m(\phi)$ acquires explicit dependence on the SSB vacuum $v(a)$, so that the effective coupling $\beta(a)$ inherits its time variation (see \cite{arcadi2019real} for Higgs-portal DM where the mediator mass tracks the scalar VEV). In the following sections, we generalize the case of a coupled dark sector with a $\mathbb{Z}_2$-symmetric quartic potential, as \(\phi\) develops a non-zero VEV.

\subsection{Symmetry Breaking and Dark Matter Mass Generation}
The scalar field $\phi$ couples with the fermionic DM field $\psi$ via a Yukawa interaction. The total scalar effective potential then receives three contributions: (i) the scalar self-potential $V(\phi)$, (ii) the Yukawa-induced term $y\,\phi\,\bar\psi\psi$, and (iii) interaction-generated (for non-relativistic CDM) corrections from the cosmological background, obtained by replacing the bilinear with its mean-field expectation value $\langle\bar\psi\psi\rangle\simeq\rho_{\rm DM}/m_\psi(\phi)$. 
Schematically,
\begin{equation}
V_{\rm eff}(\phi,a) = V(\phi) + y\,\phi\,\langle\bar\psi\psi\rangle_{\rm cosmo} + \Delta V_{\rm loop}(\phi),
\label{eq:veff_total}
\end{equation}
where the last term encodes Coleman--Weinberg radiative corrections from fermion loops \cite{coleman1973radiative}. 

At the tree level, the Yukawa piece acts as a linear source term, displacing the symmetric minima and dynamically selecting one branch of the broken phase. The leading one-loop fermion contribution is (in $\overline{\rm MS}$):
\begin{equation}
    \Delta V_{\rm 1-loop}(\phi) \;=\; -\frac{m_\psi(\phi)^4}{64\pi^2}\Big[\ln\!\frac{m_\psi(\phi)^2}{\mu^2}-\tfrac{3}{2}\Big].
\end{equation}The classical source term yields a background KG source $\propto y\langle\bar\psi\psi\rangle_{\rm cosmo}$ (equivalently $\propto \beta\,\bar\rho_{\rm DM}$ upon identifying $\beta\sim yM_{\rm Pl}/m_\psi$), while the loop term renormalises $(\mu^2,\lambda)$ and must be controlled (via small $y$ or symmetry) to preserve radiative stability of $V(\phi)$ \cite{pradosh2025loop}. Once, the scalar self-potential retains the $\mathbb{Z}_2$ form,
\begin{equation}
V(\phi) = \frac{1}{2}\,\mu_{\rm eff}^2(a)\,\phi^2 + \frac{\lambda}{4}\,\phi^4,
\label{eq:veff_scalar}
\end{equation}
where the effective mass term becomes density dependent, $\mu_{\rm eff}^2(a)=\mu^2-\xi\,\rho_{\rm DM}(a)$. The above equation is an equivalent description of Eq \eqref{eq:higgs_potential} near the transition at the broken minimum $(\mu_{\rm eff}^2=-\lambda v^2)$ and the VEV satisfies $(v^2=-\mu_{\rm eff}^2/\lambda)$. We adopt the expanded form here for clarity when discussing the density-dependent mass term and the onset of symmetry breaking.

At early epochs ($\mu_{\rm eff}^2>0$) the potential minimum lies at $\phi=0$, keeping the symmetry unbroken and the DM--DE coupling suppressed. As the background DM density $\rho_{\rm DM}$ redshifts below a critical value, $\mu_{\rm eff}^2$ changes sign at $a=a_c$, triggering symmetry breaking and driving the field toward a nonzero VEV $v(a)\neq0$. This transition marks the onset of the dark-sector coupling and the epoch when DM effectively gains mass through interaction with the scalar condensate. 

For instance, one finds that expanding around the perturbing background $\phi(x)=v(a)+\delta\phi(x)$, the Yukawa interaction given by,
\begin{equation*}
\mathcal{L}_{\rm int} = -y\,\phi\,\bar\psi\psi 
= -y\,v(a)\,\bar\psi\psi - y\,\delta\phi\,\bar\psi\psi,
\end{equation*}acts as a fermion mass contributor, while the second term on the right-hand side describes a residual mediator coupling. Because the order parameter $v(a)$ evolves with the cosmic background, the DM mass and coupling are both time dependent. Once $\phi$ develops a nonzero VEV, the interaction becomes dynamically relevant and cannot be neglected beyond the critical epoch. A quantitative demonstration of this density-triggered transition and its backreaction on $m_\psi(a)$ is provided in Appendix~\ref{app:densitydep}.

\subsection{Theoretical Viability and Microphysical Constraints}
\label{sec:dynamical}
\subsubsection{Adiabatic tracking}
\label{sec:adiabatic}
Around the minimum of the symmetry-breaking potential, the curvature is set by \begin{equation}
    V''(v)=m_\phi^2=2\lambda v^2
\end{equation} where $\lambda$ is the quartic self-coupling. For the scalar field to adiabatically follow the evolving minimum $v(a)$ without a large phase lag, its effective mass must remain much larger than the Hubble rate. This requires:
\begin{equation*}
\frac{m_\phi^2(a)}{H^2(a)}=\frac{2\lambda v(a)^2}{H^2(a)} \gg 1,
\label{eq:adiabaticity_condition}
\end{equation*}
such that at any epoch the field mass must be much larger than the Hubble rate \cite{copeland2006dynamics}.

For the numerical estimate, the simplest choice is to evaluate at the present epoch, where both \(v(a) \to v_0\) and \(H(a) \to H_0\). Taking $\lambda\simeq 10^{-1}$ and $v_0/M_{\rm Pl}\simeq 10^{-5}$ (with the reduced Planck mass $M_{\rm Pl}\simeq 2.4\times 10^{18}\,\mathrm{GeV}$), one obtains an enormous hierarchy:
\begin{equation*}
\frac{m_\phi^2}{H_0^2} \;=\; \frac{2\lambda v_0^2}{H_0^2}
\simeq 5\times 10^{109}\;\sim\;10^{110},
\end{equation*}
such that $m_\phi\gg H_0$ at late times, the field adiabatically follows the instantaneous minimum, so that the background trajectory may be identified with the VEV, $\bar\phi(a)\simeq v(a)$. This justifies treating the effective coupling as $\beta(\phi)\to\beta[v(a)]\equiv\beta(a)$ when inserted into the background exchange term $Q$ and its perturbative counterpart $\mathcal{Q}=aQ$.

\subsubsection{Radiative stability}
\label{sec:radiativestab}
 Large dark-sector VEVs and heavy mediator masses are commonly assumed in dark-Higgs and hidden-sector model building; such scales may be technically natural if protected by symmetries (such as discrete or shift symmetries, supersymmetry, or classical scale invariance) or if portal couplings to the SM are sufficiently small (see \cite{batell2011dark, heikinheimo2013twin}). 

At one-loop order, the Coleman--Weinberg effective potential \cite{camargo2016all, Morozumi2011Quantum} is:
\begin{equation}
V_{\rm eff}(\phi)=V(\phi)+\frac{1}{64\pi^2}\sum_i (-1)^{F_i}\,m_i^4(\phi)\left[\ln\!\frac{m_i^2(\phi)}{\mu^2}-c_i\right],
\label{eq:CW}
\end{equation}
where $m_i(\phi)$ are field-dependent masses, $F_i$ is the fermion number, $\mu$ is the renormalization scale, and $c_i$ are scheme-dependent constants (e.g. $3/2$ in $\overline{\rm MS}$). For Yukawa-coupled fermionic DM with $m_\psi(\phi)=m_0+y\phi$, loop corrections renormalize $(\mu^2,\lambda)$ and shift the VEV. The fractional correction is approximately:
\begin{equation*}
    \frac{\Delta v}{v}\sim \frac{y^4}{16\pi^2\lambda}.
\end{equation*}
For $\lambda\gtrsim 10^{-2}$ and $y\lesssim 10^{-2}$, typical of couplings yielding $\beta_0=\mathcal{O}(1)$ at late times (Eq.~\eqref{eq:beta_yukawa_time}), this correction is negligible ($\Delta v/v\lesssim 10^{-4}$). 

We note that the precise numerical shift depends on the choice of renormalization scale $\mu$, which is typically fixed near the physical mass threshold to minimize logarithmic running, as shown in Appendix \ref{app:renommass}. In a UV completion containing heavier dark-sector states, additional loop effects could in principle generate cubic or higher-dimensional operators (such as \ $\phi^3$ or $(\phi^2/M^2)$). Still, such terms are absent in the minimal setup considered here and can be consistently suppressed if the heavy spectrum respects the $\mathbb{Z}_2$ symmetry of the potential \cite{upadhye2013symmetron}. Additional diagnostics in the Supplementary Material confirm this scenario, with the rise of the coupling indeed following the expected order-parameter behavior, and that our logistic parametrization (see section \ref{sec:cospara}) accurately captures the underlying microphysics.

\subsubsection{Mediator-induced dark matter self-interactions}

Fluctuations $\delta\phi$ around the SSB vacuum act as light mediators that generically induce DM self-scattering. In the Born regime, the momentum-transfer cross section per unit mass for fermionic DM with Yukawa coupling $y$ can be estimated as:
\begin{equation*}
\frac{\sigma_{\rm SI}}{m_\psi}\;\sim\;\frac{y^4}{4\pi m_\psi^3}\,\frac{1}{\big(1+m_\phi^2/q^2\big)^2},
\label{eq:SI_cross_section}
\end{equation*}
where $q\sim m_\psi v_{\rm rel}$ is the momentum transfer and $v_{\rm rel}$ is the relative velocity. Astrophysical probes such as halo ellipticities and merging clusters place an approximate upper bound of $\sigma/m \lesssim 1\,{\rm cm^2\,g^{-1}}$ \cite{randall2008constraints}. For representative parameters $m_\psi\simeq 100\,{\rm GeV}$ and $y\simeq 10^{-17}$ (corresponding to $\beta_0\simeq 0.5$ from Eq.~\eqref{eq:beta_yukawa_time}), and assuming $m_\phi\gg q$, we find $\sigma/m\sim 10^{-74}\,{\rm cm^2\,g^{-1}}$, i.e. many orders of magnitude below observational limits. An analogous estimate holds for scalar-portal DM [Eq.~\eqref{eq:scalar_mass_shift}], where the couplings $g_{\chi\chi\phi}=2gv$ and $g_{\chi\chi\phi\phi}=g$ likewise yield suppressed cross sections for phenomenologically viable $g$. 

Although $yM_{\rm Pl}/m_0$ can produce an $\mathcal{O}(1)$ $\beta$, because of the large $M_{\rm Pl}$, the coupling of DM to SM particles may be zero or loop-suppressed. Direct detection experiments probe DM–SM couplings; if the mediator $\phi$ couples only feebly (or not at all) to SM, direct signals vanish.  Conversely, if $\phi$ couples to SM but is heavy, exchange is short-range and becomes hard to detect. Either way, there are parameter regions safe from current bounds.

\subsubsection{Physical scale of the interaction kernel}
\label{sec:backgroundQ}
The kernel $Q$ controls the rate of energy exchange between DM and DE at the background level. Adopting fiducial late-time values $H_0\simeq70\,{\rm km\,s^{-1}Mpc^{-1}}$, 
$\bar{\rho}_{c0}\sim10^{-6}\,{\rm GeV}^4$, and $\dot{\bar{\phi}}/M_{\rm Pl}\sim H_0$, the present-day magnitude is
\begin{equation*}
Q_0 \;\sim\;\beta_0\,\frac{\bar{\rho}_{c0}\,\dot{\bar{\phi}}}{M_{\rm Pl}} 
\;\simeq\;\beta_0\times1.5\times10^{-48}\,{\rm GeV}^5,
\label{eq:Q_magnitude}
\end{equation*}
with dimensions [energy density] $\times$ [time$^{-1}$]. The canonical dilution term is $3H_0\bar{\rho}_{c0}\simeq4.5\times10^{-48}\,{\rm GeV}^5$, so their ratio is
\begin{equation*}
\frac{Q_0}{3H_0\bar{\rho}_{c0}} \;\simeq\; \frac{\beta_0}{3}.
\end{equation*}
Thus, for $\beta_0\lesssim10^{-2}$ the background-level correction is at most $\mathcal{O}(10^{-3})$, well below current distance-measurement precision, while $\beta_0=\mathcal{O}(1)$ would induce order--unity modifications. Although $Q$ is suppressed at the background level ($Q\ll 3H \bar{\rho}_{c0}$), it still enters linearly in the continuity and Euler equations such that the leading observable effects of $\beta(a)$ arise in the growth of structure rather than in modifications to the background expansion.

\section{Cosmological Parametrization of the Coupling}
\label{sec:cospara}

In this section, we develop a compact parametrization of the coupling that is directly traceable to the microphysical SSB mechanism discussed so far. We derive an effective, epoch-dependent coupling \(\beta(a)\) and an exchange kernel \(Q(a)\) that fulfill three main criteria: (i) retain the link to the underlying scalar–fermion Lagrangian, (ii) expose the minimal set of phenomenological parameters that control amplitude, onset and sharpness of the interaction, and (iii) are convenient for numerical implementation in the background and perturbation equations in Sec. \ref{sec:observationalconstraints}.

We begin from a minimally coupled dark-sector Lagrangian
\begin{equation}
\mathcal{L} = \frac{1}{2}\partial_\mu\phi\,\partial^\mu\phi - V(\phi) 
+ \bar\psi\big(i\slashed{\nabla} - m_0 - y\phi\big)\psi,
\label{eq:L_coupled}
\end{equation}
whose variation yields the scalar Klein–Gordon equation with a Yukawa source and the Dirac equation with a field-dependent mass,
\begin{equation}
\Box\phi + V'(\phi) = y\,\bar\psi\psi,\qquad
\big(i\slashed{\nabla} - m_\psi(\phi)\big)\psi = 0,\quad m_\psi(\phi)=m_0+y\phi.
\end{equation}
Taking the cosmological expectation value of the fermion bilinear,
\(\langle\bar\psi\psi\rangle\), translates the microscopic source into an effective density-dependent term at the background level. Once the scalar acquires a nonzero VEV, this interaction manifests macroscopically as a time-dependent coupling between DM and DE. A convenient piecewise representation of the background VEV that captures this process is
\begin{equation}
v(a)=
\begin{cases}
0, & a\le a_c,\\[4pt]
\sqrt{-\mu_{\rm eff}^2(a)/\lambda}, & a>a_c,
\end{cases}
\label{eq:vev_piecewise_repeat}
\end{equation}
with \(a_c\) the critical epoch at which \(\mu_{\rm eff}^2\) changes sign. 

For the two classes of couplings discussed in the previous section, one obtains the microscopic-to-macroscopic mappings
\begin{align}
\text{(Yukawa fermion)}\quad & \beta(a)=\frac{y\,M_{\rm Pl}}{m_0+y\,v(a)}, \label{eq:beta_yuk}\\
\text{(scalar portal)}\quad & \beta(a)=\frac{M_{\rm Pl}\,g\,v(a)}{m_0^2+g\,v(a)^2} . \label{eq:beta_portal}
\end{align}
Both expressions make explicit that the epoch dependence of \(\beta(a)\) is set by the order parameter \(v(a)\).  In the regimes \(m_0\gg yv\) (Yukawa) or \(m_0^2\gg g v^2\) (portal), the coupling is approximately proportional to \(v(a)\), motivating the phenomenological sigmoid adopted below.

\subsection{Relaxation dynamics and logistic approximation}
Near the transition, the coarse-grained evolution of the order parameter is well described by an overdamped relaxational equation,
\begin{equation}
\dot v \;=\; \Gamma\big[\,\mu_{\rm eff}^2(a)\,v - \lambda\,v^3\,\big],
\label{eq:dvdt_basic_repeat}
\end{equation}
with \(\Gamma>0\) an effective relaxation rate.  Switching to e-fold time \(N=\ln a\) (so \(\dot v=H\,dv/dN\)) and defining the dimensionless coefficients
\begin{equation}
s(a)\equiv\frac{\Gamma\,\mu_{\rm eff}^2(a)}{H(a)},\qquad
\tilde\lambda(a)\equiv\frac{\Gamma\lambda}{H(a)},
\end{equation}
Eq.~\eqref{eq:dvdt_basic_repeat} becomes
\begin{equation}
\frac{dv}{dN}=s(a)\,v-\tilde\lambda(a)\,v^3.
\label{eq:dv_dN}
\end{equation}
For a narrow transition window, it is justified to approximate \(s(a)\simeq s\) and \(\tilde\lambda(a)\simeq\tilde\lambda\) as slowly varying, in which case the rescaled variable \(w\equiv v/v_\infty\) (with \(v_\infty\equiv\sqrt{s/\tilde\lambda}\)) obeys
\begin{equation}
\frac{dw}{dN}=s\,w(1-w^2).
\end{equation}
The logistic replacement \(1-w^2\to 1-w\) captures the leading sigmoidal growth while simplifying the mapping from microphysics to phenomenology; its closed form solution with the midpoint fixed at \(a=a_c\) reads
\begin{equation}
w(a)=\frac{a^s}{a^s+a_c^s}\quad\Longrightarrow\quad
v(a)=v_\infty\frac{a^s}{a^s+a_c^s}.
\label{eq:v_sigmoid_repeat}
\end{equation}
Identifying \(n\equiv s\) yields the commonly used algebraic sigmoid parametrization
\begin{equation}
\beta(a)\simeq\beta_0\frac{a^n}{a^n+a_c^n},
\label{eq:beta_param_repeat}
\end{equation}
where \(\beta_0\) denotes the late-time plateau of the coupling (set by the microscopic parameters \(y,m_0,v_\infty\) or \(g,m_0,v_\infty\) through \eqref{eq:beta_yuk}–\eqref{eq:beta_portal}). Alternative transition shapes—such as exponential or hyperbolic-tangent functions—can in principle reproduce similar behavior to that of Eq~\eqref{eq:beta_param_repeat}, but we choose the logistic form adopted here as the simplest analytic mapping to the underlying microphysics and the most numerically stable implementation for cosmological inference.

\subsection{Parametrized evolution of $v(a)$ and $Q(a)$}
To capture the smooth activation of the symmetry-breaking phase transition in a compact form, it is useful to introduce a normalized activation function
\begin{equation}
f(a)\equiv\frac{a^n}{a^n+a_c^n},\qquad v(a)=v_\infty f(a),\qquad \beta(a)=\beta_0 f(a).
\end{equation}
Using \(d/dt=H\,d/dN\), we obtain the explicit time derivative
\begin{equation}
\dot v(a)=H(a)\,n\,v_\infty\,f(a)\big[1-f(a)\big],
\label{eq:v_dot}
\end{equation}
and substituting into the conformal exchange kernel \(Q(a)=(\beta(a)/M_{\rm Pl})\bar\rho_{\rm DM}\,\dot v(a)\) gives the compact diagnostic form
\begin{equation}
Q(a)=\frac{\beta_0}{M_{\rm Pl}}\,n\,v_\infty\,H(a)\,\bar\rho_{\rm DM}(a)\; f^2(a)\big[1-f(a)\big].
\label{eq:Q_compact}
\end{equation}
Equation~\eqref{eq:Q_compact} is analytically equivalent to the expression used in our numerical solver but exposes the physical structure of the interaction term—its scaling with the background expansion $(H(a))$, the DM density $(\bar\rho_{\rm DM}(a))$, and the activation kernel $(f(a))$. 

\begin{figure*}
    \centering
    \includegraphics[width=0.65\linewidth]{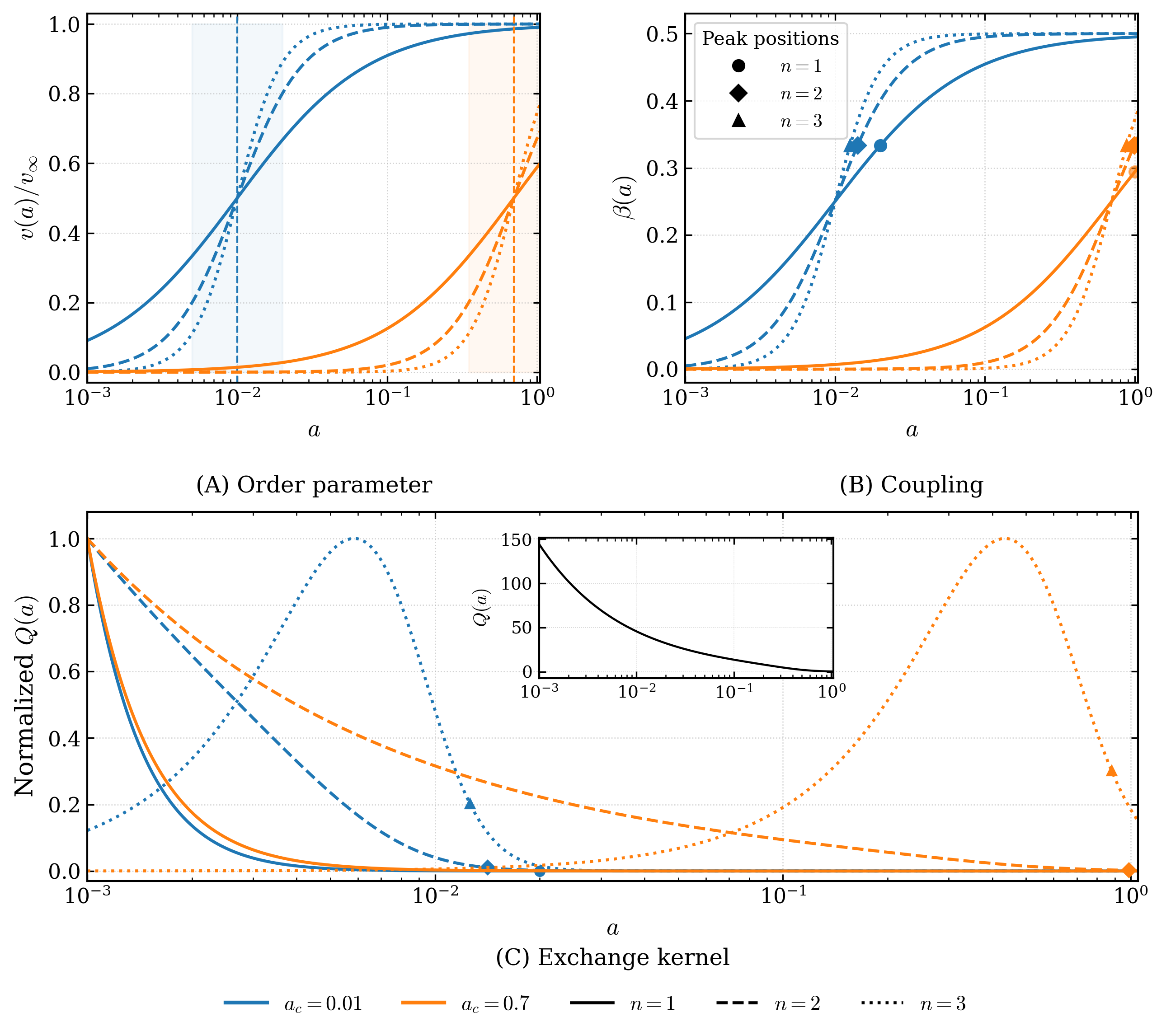}
    \caption{ \small Panel~(A) shows the normalized order parameter $v(a)/v_\infty$, which rises smoothly from the symmetric phase ($v=0$) to its asymptotic value as the effective mass term changes sign. Panel~(B) displays the corresponding coupling $\beta(a)=\beta_0\,f(a)$, whose evolution directly follows that of $v(a)$ through Eq.~\eqref{eq:v_dot}. Panel~(C) shows the normalized kernel $Q(a)$, peaking shortly after the transition epoch when the field’s relaxation rate $\dot v$ is maximal. The inset highlights the background Hubble scaling factor in Eq.~\eqref{eq:Q_compact}, showing that the overall amplitude of $Q(a)$ is dominated by the late-time acceleration regime.
}

    \label{fig:ssbfig}
\end{figure*}

As illustrated in Fig.~\ref{fig:ssbfig}, the kernel $Q(a)$ peaks shortly after the transition epoch when the relaxation rate $\dot{v}$ is maximal. For $n=1$ and $a_c=0.7$, the peak occurs slightly after the present epoch ($a>1$), indicating that the phase transition is still ongoing today. For $n=2$, the peak aligns approximately with the present epoch ($a=1$), while for $n \gtrsim 3$, the activation is sharply localized around $a \simeq a_c$, corresponding to an abrupt onset of the coupling. This shows that the piecewise construction enforces a strictly symmetric phase ($v=0$) for $a \leq a_c$, while the sigmoid form in Eq.~\eqref{eq:v_sigmoid_repeat} provides a smooth transition for moderate $n$. The resulting exchange kernel from Eq.~\eqref{eq:Q_compact} vanishes in the unbroken phase and peaks shortly after $a_c$, as the coupling induces a finite, transient energy-momentum transfer between the dark components. Thus, the parameters $a_c$ and $n$ directly control the onset and sharpness of this activation, respectively. Note that this behavior also captures the degree of abruptness in the transition of the order parameter between phases, guiding the practical parameter choices adopted in subsequent sections.

\subsection{Implementation notes, assumptions, and practical choices}

We now summarize the practical setup used for the observational analysis, including the parameter set varied in the MCMC runs, the adopted prior ranges, and the key approximations entering the implementation. The goal is to make explicit the assumptions that link the microscopical model parameters introduced in Sec.~\ref{sec:microphysics} with the coupling description entering the data analysis.

\medskip
\noindent\textbf{Parameter set:}
We vary the parameters \(\{\Omega_m, H_0, \sigma_{8,0}, \beta_0, a_c, n\}\),
where \(\beta_0\) controls the present-day coupling strength,
\(a_c\) denotes the activation epoch of the interaction,
and \(n\) sets the sharpness of the transition.
Depending on the adopted microphysical prescription, we either
(i) fix \(v_\infty\) through the mappings in Eqs.~\eqref{eq:beta_yuk}--\eqref{eq:beta_portal}, 
or (ii) absorb it into the normalization of \(\beta_0\).
These correspond respectively to a microphysically anchored model
and a purely phenomenological effective description.
Each value of \(a_c\) defines a distinct realization of the model,
representing early-, mid-, or late-time activation of the coupling.

\medskip
\noindent\textbf{Priors:}
To explore the phenomenologically relevant region, we adopt the following prior ranges:
\[
\beta_0 \in [0,2]\ \text{(uniform)},\qquad
a_c \in [10^{-3},1]\ \text{(log-uniform for }a_c\ll1),\qquad
n \in [1,4]\ \text{(uniform)}.
\]
These intervals encompass weak to moderately strong couplings and phase transitions ranging from gradual to sharp across a wide span of epochs. They are sufficiently broad to include all viable posteriors identified in exploratory runs. Specific choices and convergence diagnostics for each MCMC chain
are reported in Sec.~\ref{sec:observationalconstraints}.

\medskip
\noindent\textbf{Principal approximations:}
Our analysis adopts several controlled approximations, summarized below, together with their physical justification:
\begin{enumerate}
  \item \emph{Logistic approximation.}  
  The replacement \(1-w^2 \rightarrow 1-w\) reproduces the qualitative sigmoid behavior of the transition and simplifies the mapping between the microscopic slope parameter \(s\) and the phenomenological steepness index \(n\).  
  This is accurate provided the transition remains overdamped and not ultra-violent, such that the cubic saturation term stays subdominant.
  \item \emph{Adiabatic background.}  
  The background expansion rate is fixed to the fiducial flat \(\Lambda\)CDM form when computing growth functions (see Sec.~\ref{sec:adiabatic}).  
  This is justified because the fractional correction \(\Delta H/H\) remains well below unity over the entire posterior range (see Supplementary Material II).
  \item \emph{Microphysical matching.}  
  In parameter scans, \(v_\infty\) is numerically absorbed into \(\beta_0\), treating the latter as the effective late-time coupling.  
  For physically motivated priors, one may alternatively map \(\beta_0\) to the microscopic parameters \((y,m_0,v_\infty)\) or \((g,m_0,v_\infty)\) and impose the constraints discussed in Appendix~\ref{app:renommass}.
\end{enumerate}

\medskip
For a smooth comparison between phenomenological and microscopic representations, we recommend reporting results both in the \(\{\beta_0,a_c,n\}\) basis and in terms of the underlying Lagrangian parameters. This dual approach allows for a direct interpretation of cosmological constraints in relation to particle physics quantities, which may be of interest to readers. However, for our primary analysis of the model, we will report only in terms of model parameters.

\section{Observational Constraints on Epoch-Dependent Coupling}
\label{sec:observationalconstraints}
We begin this section by outlining the methodology and datasets used to obtain the observational constraints reported below. Our analysis primarily relies on late-time probes that are sensitive to both the cosmic expansion history and the growth of structure. Specifically, we employ redshift-space distortion (RSD) measurements of $f\sigma_8(z)$ following the compilation of Ref.~\cite{kazantzidis2018evolution}, baryon acoustic oscillation (BAO), and cosmic chronometer (CC) data to constrain $H(z)$, and the Pantheon+SH0ES supernova sample to anchor the luminosity distance at low redshift.  For the CC dataset, we adopt the 31 $H(z)$ measurements compiled from Refs.~\cite{jimenez2023cosmic, verde2024tale, d2023cosmographic, moresco20166, Mishra:2025vpy}, spanning the range $0.1<z<2$ and obtained via the differential-age (cosmic chronometer) method.  In the BAO sector, we use the 26 non-correlated line-of-sight measurements from Refs.~\cite{delubac2015baryon, di2016curvature, blake2012wigglez, chuang2013clustering}, 
which provide independent constraints on the late-time expansion rate.  For supernovae, we employ the Pantheon+SH0ES compilation \cite{riess2022comprehensive, brout2022pantheon+, brout2022panth+, scolnic2022pantheon+, Kavya:2024ssu, oka2014simultaneous}, which delivers the most precise relative distance moduli and a robust local calibration of $H_0$. 

The complete likelihood analysis is performed under two distinct background treatments:  (i) a fixed $\Lambda$CDM background, in which the Hubble rate $H(a)$ follows the standard Friedmann solution and the coupling affects only perturbations; and  (ii) a dynamically evolved background, where $H(a)$ and the scalar field are solved self-consistently according to Eqs.~(\ref{eq:v_dot})–(\ref{eq:Q_compact}).  This allows us to quantify the impact of backreaction from the coupling on both the expansion history and the growth of structure.  In both cases, we employ MCMC sampling to infer the posterior distributions of the cosmological and coupling parameters.

\subsection{Methodology: growth equations and observables}
\label{sec:methodology}

In the standard $\Lambda$CDM model, the growth of the linear matter overdensity $\delta_m \equiv \delta\rho_m/\bar{\rho}_m$ obeys~\cite{nesseris2017tension}:
\begin{equation}
\ddot{\delta}_m + 2H \dot{\delta}_m - 4\pi G \,\bar{\rho}_m\, \delta_m = 0 .
\label{eq:standard_growth}
\end{equation} In IDE scenarios, however, the interaction term $Q(a)$ alters the continuity and Euler equations (see Eqs.~\eqref{eq:KG_perturbed} -- \eqref{eq:de_euler_corrected_conf} ), while the scalar mediates an effective fifth force between DM particles. In the small-field limit, where the scalar is effectively massless on linear cosmological scales and screening effects are negligible, the modified Newton’s constant takes the simple form~\cite{amendola2004phantom, amendola2006constraints, yoo2012theoretical}:
\begin{equation}
G_{\mathrm{eff}}(a) = G \left[ 1 + 2\,\beta^2(a) \right],
\label{eq:6.2}
\end{equation}
for which the field is adiabatic at the background level (suppressing $Q$) while still mediating an unsuppressed fifth force on sub-horizon modes. The growth equation then becomes:
\begin{equation}
\ddot{\delta}_m + \left[ 2H + \frac{Q}{\bar{\rho}_m} \right] \dot{\delta}_m
  = 4\pi G_{\mathrm{eff}}(a)\,\bar{\rho}_m\,\delta_m .
\label{eq:modified_growth}
\end{equation}

For numerical integration, we rewrite Eq.~\eqref{eq:modified_growth} in terms of the e-fold variable $N \equiv \ln a$, using $d/dt = H\,d/dN$. Denoting \( '\equiv d/dN\), one obtains:
\begin{equation}
\delta_m'' + \left[ 2 + \frac{H'}{H} - \frac{\beta(a)}{M_{\mathrm{Pl}}}\frac{\dot{\bar\phi}}{H} \right]\delta_m'
   = \frac{3}{2}\,\Omega_m(a)\left[1+2\beta^2(a)\right]\delta_m ,
\label{eq:delta_evolution}
\end{equation}
where $\Omega_m(a) \equiv \bar{\rho}_m / (3 M_{\mathrm{Pl}}^2 H^2)$. Equivalently, we rewrite the above expression as:
\begin{equation}
\delta_m'' + \left[ 2 + \frac{d\ln H}{d\ln a} + Q_{\rm fric}(a) \right]\delta_m'
   - \frac{3}{2}\,\Omega_m(a)\left[1+2\beta^2(a)\right]\delta_m = 0 ,\label{eq:deltaequi}
\end{equation}
with $Q_{\rm fric}(a)\equiv Q/(H\bar\rho_m)$ representing the effective friction induced by the interaction. The two representations \eqref{eq:delta_evolution} and \eqref{eq:deltaequi} are algebraically equivalent once \(Q(a)\) is expressed in terms of \(\beta(a)\) and the background scalar velocity \(\dot{\bar\phi}\). This means that the early-time background dynamics reproduce the standard matter-dominated behavior, while allowing late-time deviations that can impact both the expansion history and the growth of structure.

In our numerical analysis, we consider both the adiabatic (frozen-background) approximation and the fully dynamical background evolution. In the adiabatic case, the Hubble parameter is fixed to the standard $\Lambda$CDM form, effectively treating the field as frozen until symmetry breaking and isolating the late-time impact of the coupling on structure growth:
\begin{equation}
H^2(a) = H_0^2 \left[ \Omega_m a^{-3} + (1-\Omega_m) \right].
\end{equation}
In this limit, the epoch-dependent coupling $\beta(a)$ affects only the perturbation sector, entering the modified growth equation through Eq.~\eqref{eq:6.2} and the friction term in Eq.~\eqref{eq:beta_param_repeat}.  

For the dynamical (self-consistent) background case, the expansion history is obtained by solving the coupled system,
\begin{equation}
H^2 = \frac{1}{3M_{\mathrm{Pl}}^2}\big(\rho_b + \rho_r + \rho_{\rm DM} + \rho_\phi\big), \qquad
\dot{\rho}_{\rm DM} + 3H\rho_{\rm DM} = +Q, \qquad
\dot{\rho}_\phi + 3H(\rho_\phi + p_\phi) = -Q,
\label{eq:self_consistent_background}
\end{equation}
where $\rho_\phi$ and $p_\phi$ are the scalar-field energy density and pressure, respectively. Using dimensionless variables $\tilde{\rho}_i=\rho_i/\rho_{\mathrm{crit},0}$, $\varphi=\phi/M_{\mathrm{Pl}}$, and $u=d\varphi/dN=\dot{\phi}/(H M_{\mathrm{Pl}})$, the evolution equations take the compact form
\begin{align}
\tilde{\rho}_{\rm DM}' &= -3\tilde{\rho}_{\rm DM} + \beta(a)\,u\,\tilde{\rho}_{\rm DM}, \nonumber\\[3pt]
u' &= -\!\left(3+\frac{d\ln H}{dN}\right)u
      -\frac{1}{H^2}\frac{\partial \tilde{V}(\varphi)}{\partial\varphi}
      + \beta(a)\,\frac{\tilde{\rho}_{\rm DM}}{H^2}, \nonumber\\[3pt]
\frac{H^2}{H_0^2} &= \tilde{\rho}_b + \tilde{\rho}_r + \tilde{\rho}_{\rm DM} 
                     + \tfrac{1}{2}u^2 + \tilde{V}(\varphi),
\label{eq:H_self_consistent}
\end{align}
where primes denote derivatives with respect to $N=\ln a$ and $\tilde{V}=V/\rho_{\mathrm{crit},0}$.  

The microphysical parameters $\{\lambda, v_0, \xi_{\mathrm{density}}\}$ enter the background through Eq.~\eqref{eq:higgs_potential}, contributing to $\rho_\phi$ and $V_{,\phi}$ in Eq.~\eqref{eq:KG_background_corrected}. Their effects propagate into Eq.~\eqref{eq:H_self_consistent} via the term $\tilde{V}(\varphi)$ and its slope $\partial\tilde{V}/\partial\varphi$, which govern the evolution of the scalar velocity $u=d\varphi/dN$ and hence the Hubble rate $H(a)$. The parameters $\lambda$ and $v_0$ determine the steepness and minimum of the quartic potential, while $\xi_{\mathrm{density}}$ couples the scalar amplitude to the matter density $\rho_{\rm DM}$, producing an environment-dependent effective mass $m_\phi^2=V_{,\phi\phi}(\phi=v_0)$ (see Appendix~\ref{app:densitydep}).

At each MCMC step, the background system~\eqref{eq:H_self_consistent} is first integrated from $a_{\mathrm{ini}}=10^{-5}$ to $a=1$ using a stiff \texttt{Radau} solver with relative and absolute tolerances of $10^{-9}$–$10^{-12}$ and initial conditions $\varphi(a_{\mathrm{ini}})=0$, $u(a_{\mathrm{ini}})=10^{-8}$, ensuring the field remains frozen during the radiation era. The resulting $H(a)$, $\Omega_m(a)$, and $\dot{\bar{\phi}}(a)$ are then supplied to the linear growth equation for the numerical evolution of the matter perturbation $\delta_m(a)$, solved using \texttt{solve\_ivp} with adaptive Runge–Kutta integration and tolerances $10^{-9}$ and $10^{-10}$. The growing-mode initial condition $\delta_m \propto a$ is imposed at early times ($a \ll a_c$), where the coupling is negligible and the solution reduces to the standard $\Lambda$CDM limit. The growth factor is defined as
\begin{equation}
D(a) \equiv \frac{\delta_m(a)}{\delta_m(a_{\mathrm{ini}})},
\end{equation}
and normalized to $D(a{=}1)=1$. From this, we compute the logarithmic growth rate and RSD observable:
\begin{align}
f(a) &= \frac{d\ln D}{d\ln a}, \nonumber\\
\sigma_8(z) &= \sigma_{8,0}\,\frac{D(z)}{D(0)}, \\
f\sigma_8(z) &= f(z)\,\sigma_8(z), \nonumber
\label{eq:fs8}
\end{align}
where $\sigma_{8,0}$ is fixed by the present-day normalization.

To obtain the matter power spectrum, we rescale the \(\Lambda\)CDM linear spectrum \(P_{\Lambda\mathrm{CDM}}(k)\) from \texttt{CAMB} by the square of the growth ratio,
\begin{equation}
P_\beta(k) \simeq P_{\Lambda\mathrm{CDM}}(k)\left[\frac{D_\beta}{D_{\Lambda\mathrm{CDM}}}\right]^2 ,
\label{eq:6.10}
\end{equation}
which is valid in the linear, scale-independent regime (\(k\lesssim 0.1\,h\,\mathrm{Mpc}^{-1}\)) where the modified gravitational coupling can be approximated by Eq.~\eqref{eq:6.2}. This quasi-static approximation assumes that the scalar field is effectively massless on the relevant scales (\(m_\phi a/k \ll 1\)) and that time derivatives of the field perturbation are negligible relative to spatial gradients (\(k/a\gg H\)). 

To verify that this regime applies to our parameter space, we introduce the diagnostic quantity
\begin{equation}
\mathcal{S}(k,a)\equiv\frac{m_\phi(a)\,a}{k},
\end{equation}
which measures the ratio of the Compton wavelength of the mediator to the perturbation scale. The linear, scale-independent limit corresponds to \(\mathcal{S}\ll1\), whereas \(\mathcal{S}\gg1\) indicates a Yukawa-suppressed force. For the microphysical benchmark region explored in Sec \ref{sec:results}, we find that \(\mathcal{S}\ll1\) before activation (when \(\beta\simeq0\)) and \(\mathcal{S}\gg1\) after symmetry breaking, so the intermediate regime \(\mathcal{S}\sim\mathcal{O}(1)\)—which would yield scale-dependent growth—is not realized. Hence, \(G_{\mathrm{eff}}\) remains approximately scale independent across the wavenumber range \(k\lesssim0.1\,h\,\mathrm{Mpc}^{-1}\) relevant to the late-time RSD and BAO observables. The full evolution of \(\mathcal{S}(k,a)\) and its verification across the posterior samples are shown in the \textit{Supplementary Material II}, confirming the validity of Eqs.~\eqref{eq:6.2} and \eqref{eq:6.10} within the parameter ranges considered.

\begin{figure*}
    \centering
    \includegraphics[width=0.75\linewidth]{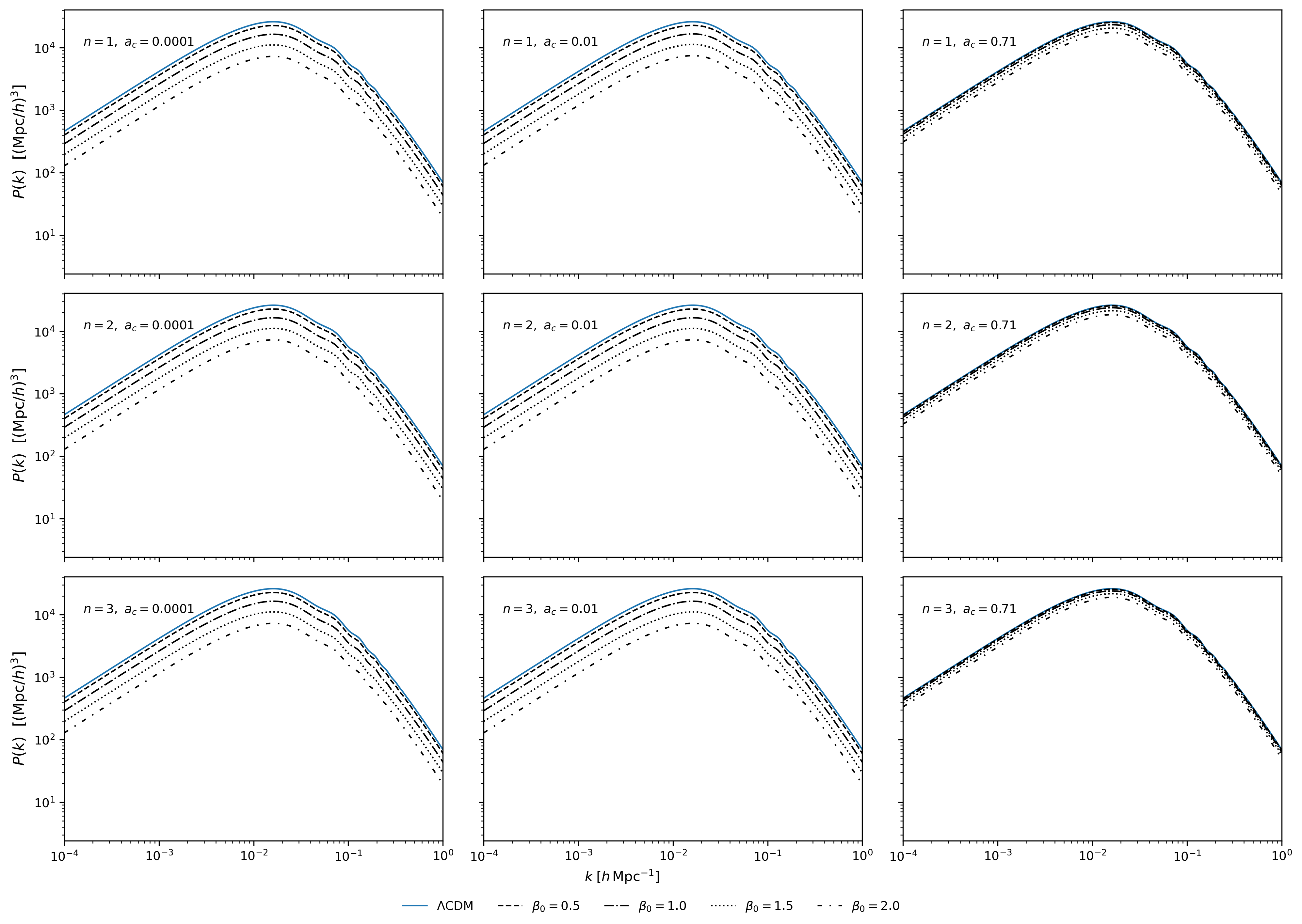}
    \caption{\small Linear matter power spectra \(P(k)\) for the fiducial \(\Lambda\)CDM model (blue solid line) and IDE models (black lines). Rows correspond to different steepness parameters \(n = 1, 2, 3\), while columns correspond to coupling epochs \(a_c = 10^{-4}\) (early radiation era), \(0.01\) (mid-matter era), and \(0.71\) (late-time). Different line styles correspond to varying present-day coupling strengths \(\beta_0\). Early-time couplings (\(a_c \ll 1\)) strongly amplify small- and intermediate-scale power, conflicting with \textit{Planck} and BAO constraints, whereas mid-to-late couplings lead to moderate growth enhancements. Late-onset of couplings (\(a_c \gtrsim 0.7\)) leaves the matter power spectrum nearly indistinguishable from the \(\Lambda\)CDM prediction. The curves represent linear-theory predictions based on CAMB transfer functions rescaled by the corresponding growth factors, intended as illustrative trends rather than fully self-consistent Boltzmann solutions. }

 \label{fig:pkvariation}
\end{figure*}

Figure~\ref{fig:pkvariation} illustrates representative \(P(k)\) curves computed from Eq.~\eqref{eq:6.10} and qualitative summary of how \((a_c,n)\) affects the growth history and \(P(k)\) is presented in Table~\ref{tab:coupling_effects_vertical}. Increasing the coupling amplitude \(\beta_0\) uniformly boosts the power across linear scales, with the largest fractional deviation at the largest modes (\(k\lesssim0.01\,h\,\mathrm{Mpc}^{-1}\)). Early-time couplings (\(a_c\ll1\)) induce excessive, often scale-dependent growth inconsistent with \textit{Planck} and BAO data—signalling the breakdown of the scale-independent approximation—whereas mid-epoch couplings (\(a_c\simeq0.01\)) yield moderate, approximately scale-independent enhancements consistent with Eq.~\eqref{eq:6.10}. Late-onset couplings (\(a_c\gtrsim0.7\)) activate only after most structure formation and remain nearly indistinguishable from \(\Lambda\)CDM on linear scales. This motivates the focus of our analysis on mid-to-late activation epochs in the main parameter scans carried out in the next section.

\begin{table*}[ht]
\scriptsize
\centering
\caption{\small Qualitative effect of epoch-dependent coupling on the matter power spectrum $P(k)$ for different $n$ and $a_c$ values. Comments are based on consistency with current CMB (\textit{Planck}), BAO, and LSS datasets.}
\label{tab:coupling_effects_vertical}

\begin{tabular}{cccp{3cm}p{3cm}}
\hline
\textbf{Steepness $n$} & \textbf{ Critical epoch $a_c$} & \textbf{Turn-on shape} & \textbf{Impact on $P(k)$ }& \textbf{Comments} \\
\hline
\multirow{3}{*}{1} 
 & $10^{-4}$ (radiation era) & Smooth, gradual & Large {boost} on all scales; growth starts early $\rightarrow$ big deviation from $\Lambda$CDM & Strongly disfavoured by \textit{Planck} CMB and high-$z$ clustering. \\
 & $0.01$ (mid-matter era) & Smooth, gradual & Moderate boost in small/intermediate scales; tilt at $k>0.01\,h\,\mathrm{Mpc}^{-1}$ & Mild tension with \textit{Planck} + BOSS if boost $>10\%$, but can alleviate $S_8$ tension. \\
 & $\sim 0.7$ (late-time) & Smooth & Minimal change to $P(k)$; nearly overlaps $\Lambda$CDM. & Fully consistent with CMB and BAO; effects too small to detect in current LSS data. \\
\hline
\multirow{3}{*}{2--3} 
 & $10^{-4}$ (radiation era) & Faster rise & Strong early-time boost; more step-like near $a_c$ & Ruled out — early deviations spoil CMB acoustic peaks. \\
 & $0.01$ (mid-matter era) & Fast/step-like & Boost mainly at small scales; large scales almost unaffected & Some parameter space allowed; potential fit to RSD $f\sigma_8(z)$ anomalies. \\
 & $\sim 0.7$ (late-time) & Very sharp & Negligible $P(k)$ change except tiny late-time growth & Observationally safe but degenerate with $\Lambda$CDM. \\
\hline
\end{tabular}

\label{tab:pkvariation}
\end{table*}

\subsection{Observational constraints and parameter estimation}
\label{sec:results}
In the transition regime \(a\sim a_c\), the epoch-dependent coupling slows the dilution of CDM and enhances the scalar-field dynamics. At the level of the full interacting theory, this feedback modifies the Hubble rate \(H(z)\) and the effective equation of state \(w_{\mathrm{eff}}(a)\). For sufficiently large coupling amplitude \(\beta_0\) and steepness \(n\), \(w_{\mathrm{eff}}\) can cross the acceleration threshold (\(w_{\mathrm{eff}}<-1/3\)) and, in extreme cases, transiently enter a phantom-like regime (\(w_{\mathrm{eff}}<-1\)), consistent with phenomenological reconstructions of \(w(z)\) \cite{capozziello2019extended, raveri2020reconstructing, di2021dark}. Here, we focus instead on observational signatures at late times, where the coupling primarily affects the linear growth factor \(D(a)\) and the growth rate \(f(a)\). Our analysis proceeds in two stages. First, we perform an RSD-only MCMC varying \(\{\Omega_m,H_0,\sigma_{8,0},\beta_0,a_c\}\) with fixed \(n=1\). Second, we extend the study to a joint likelihood including Pantheon+SH0ES, BAO, cosmic chronometers, and compressed \textit{Planck} distance priors, and compare results obtained (i) with a fixed \(\Lambda\)CDM background and (ii) when the background expansion is evolved self-consistently together with the scalar field and the coupling. Note that the \textit{Planck} priors act as strong constraints on the early-time geometry and therefore on any model modifications that feed back onto the background. Although they effectively fix the angular scale of the sound horizon and the baryon density while leaving late-time growth observables free to vary, when \(\beta(a)\) substantially affects recombination or the pre-recombination expansion, a full \textit{Planck} likelihood is required.

\begin{figure*}
    \centering
    \includegraphics[width=0.7\linewidth]{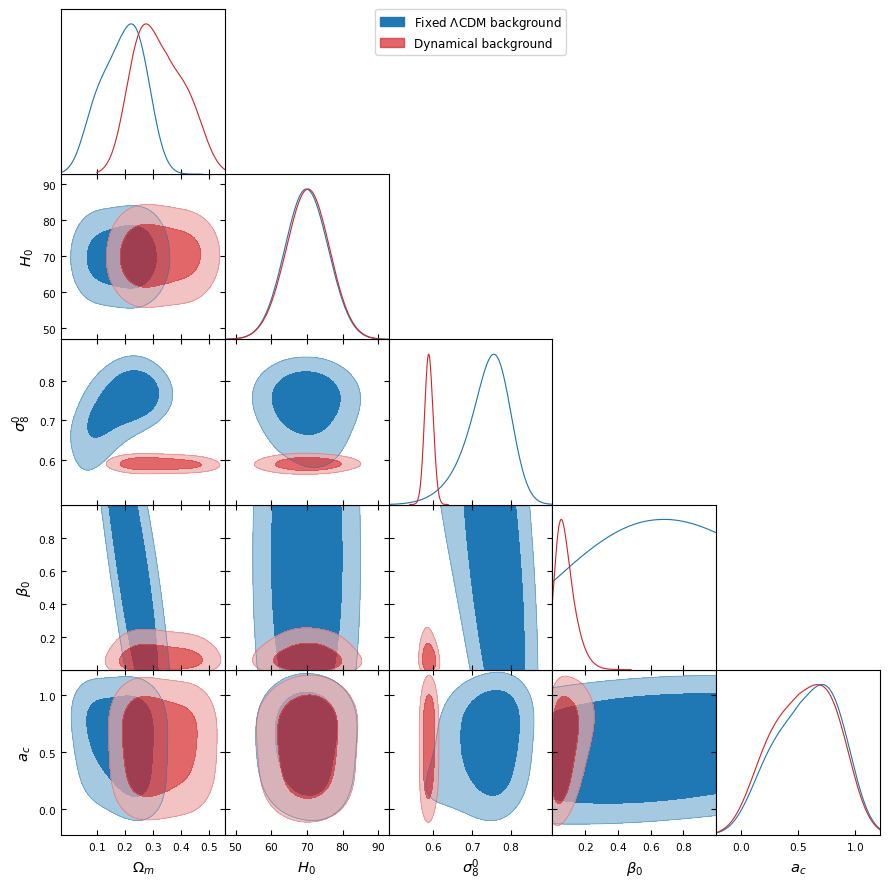}
    \caption{\small Posterior constraints on $\Omega_m$, $H_0$, $\sigma_{8,0}$, $\beta_0$, and $a_c$ from the RSD-only MCMC analysis of the epoch-dependent coupling model with fixed $n=1$. The blue contours correspond to the analysis performed with a fixed $\Lambda$CDM background, while the red contours show the results obtained when the background expansion is evolved self-consistently with the scalar–matter coupling. The dynamical background leads to a higher matter density ($\Omega_m \simeq 0.31$) and a lower growth amplitude ($\sigma_{8,0} \simeq 0.59$) compared to the fixed-background case ($\sigma_{8,0} \simeq 0.75$), indicating suppressed late-time structure formation. 
 }
    \label{fig:rsdonly}
\end{figure*}

\begin{figure*}
    \centering
    \includegraphics[width=0.32\textwidth]{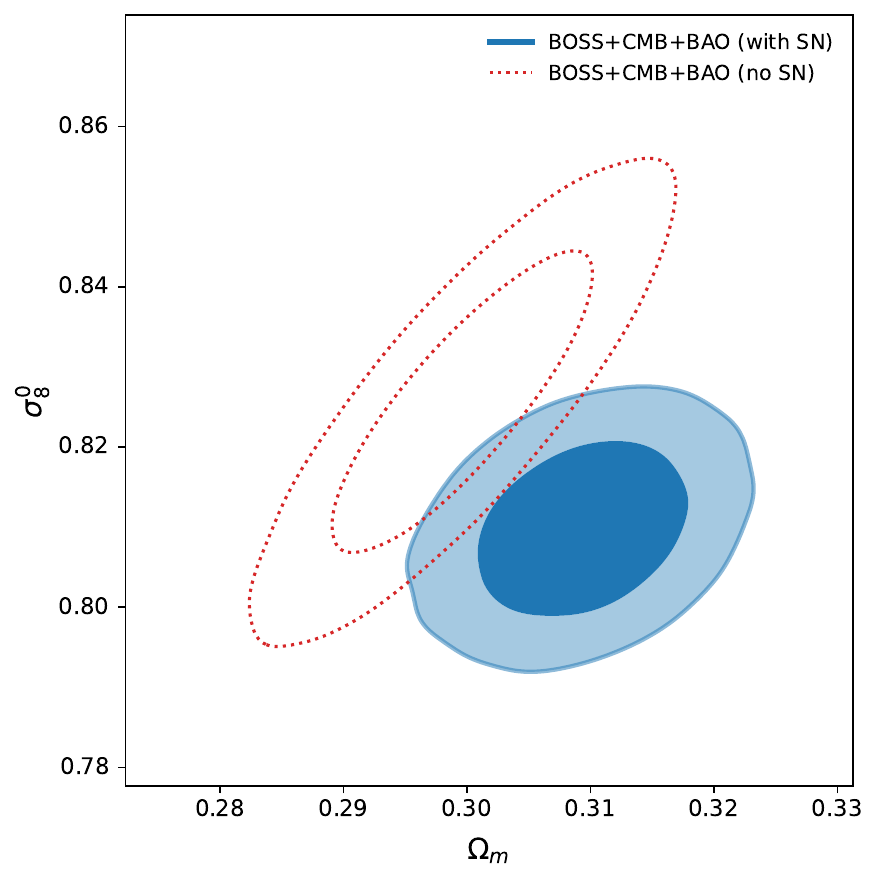}
    \includegraphics[width=0.32\textwidth]{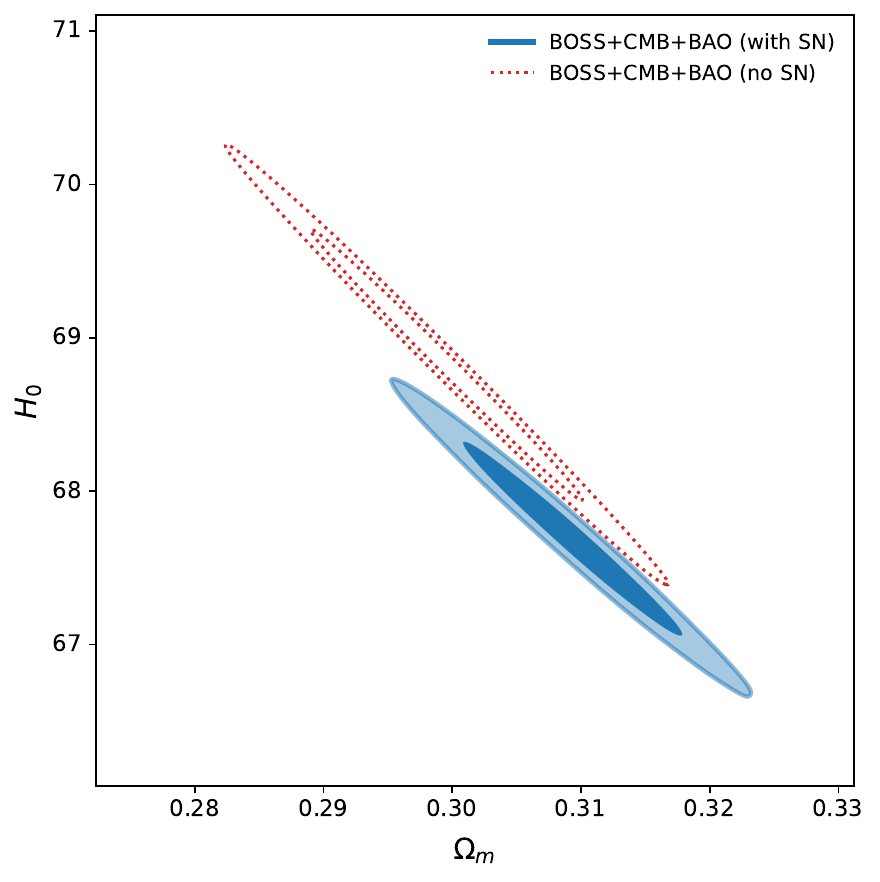}
    \includegraphics[width=0.32\textwidth]{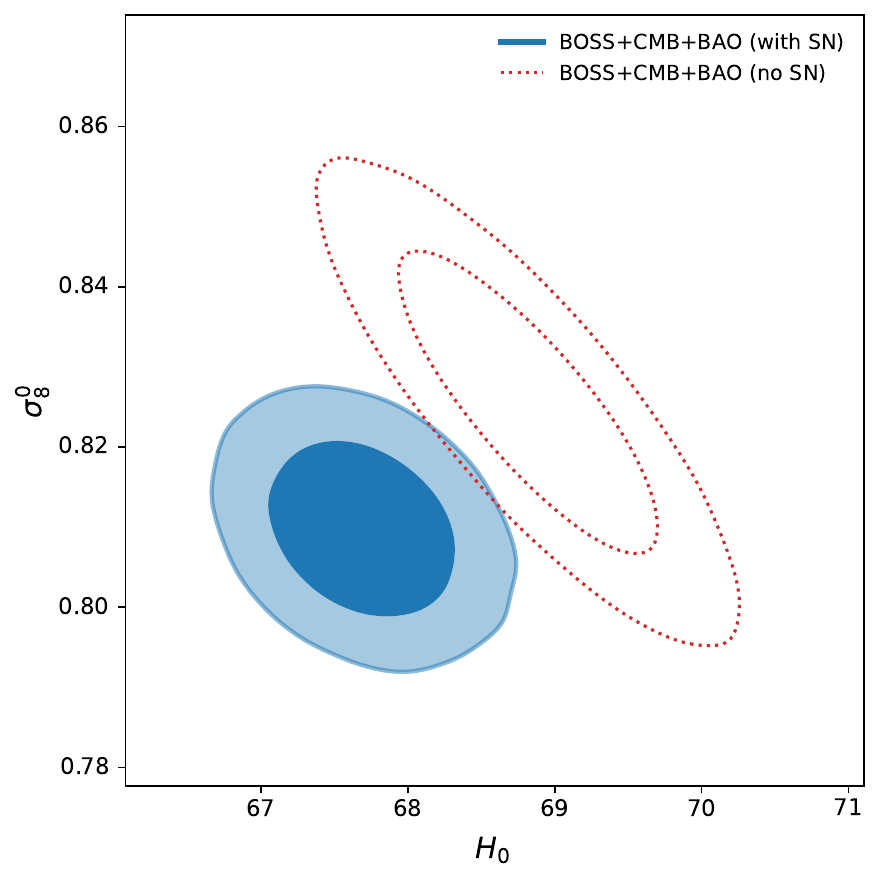}
    \caption{\small Two-dimensional marginalized posteriors with fixed \(\Lambda\)CDM background comparing the full joint analysis (BOSS RSD + \textit{Planck} priors + BAO + SN; blue, filled) with the reduced dataset (BOSS RSD + \textit{Planck} priors + BAO; red, dashed). Contours show the 68\% and 95\% credible regions. The inclusion of SN sharpens the constraints, reducing degeneracies between $\Omega_m$ and $H_0$, and slightly shifts the preferred region for $\sigma_{8,0}$ toward lower values.}
    \label{fig:two_param_posteriors}
\end{figure*}

\begin{figure*}
    \centering
    \includegraphics[width=0.6\linewidth]{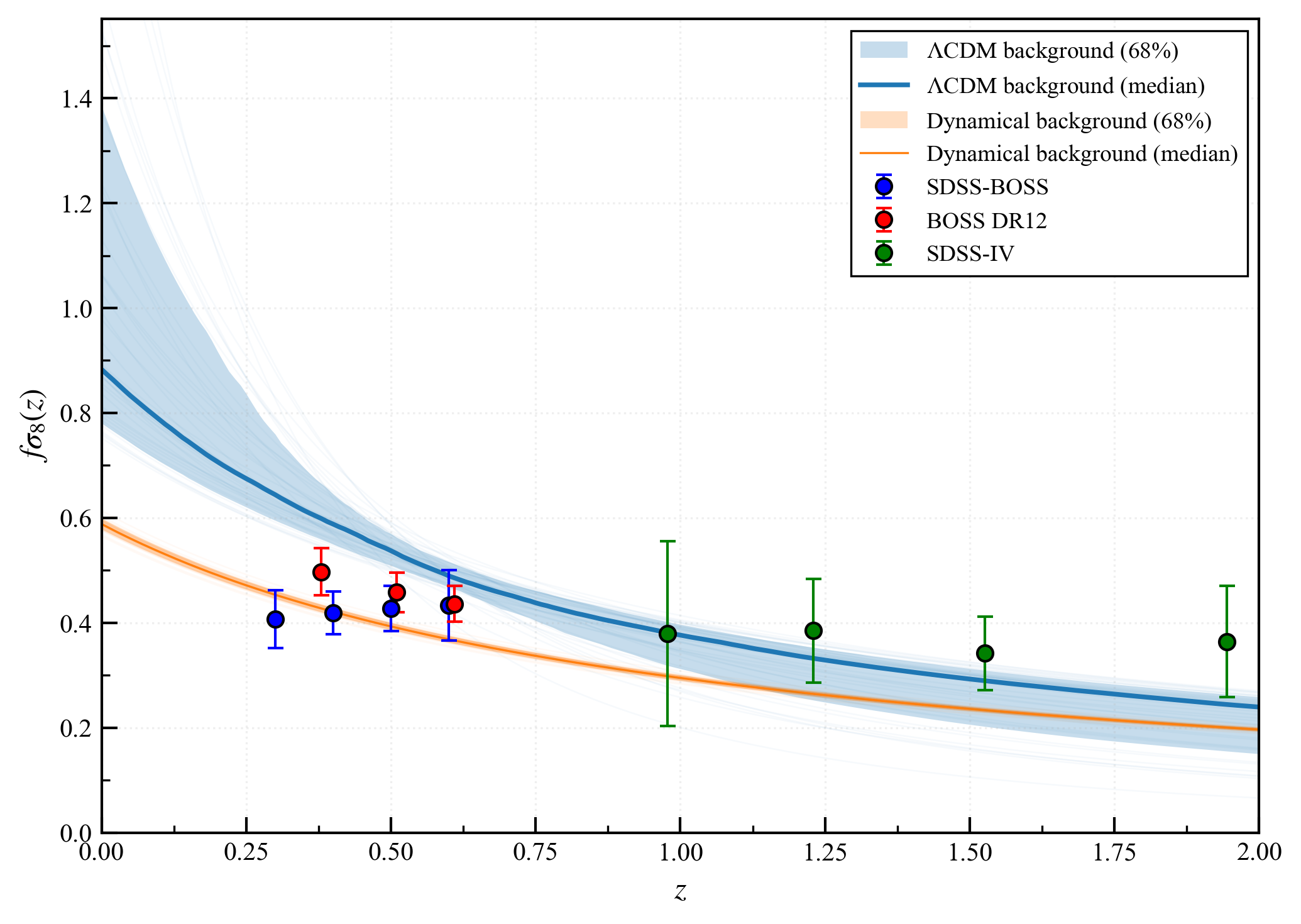}
    \caption{\small Posterior predictive distributions for the linear growth rate observable $f\sigma_8(z)$ derived from the RSD-only MCMC analysis of the model with fixed $n=1$. The solid curves represent the posterior median predictions, while the shaded regions denote the 68\% credible intervals obtained from 400 posterior samples. The blue and orange curves correspond to analyses performed with a fixed $\Lambda$CDM background and a self-consistently evolved dynamical background, respectively. The data points with $1\sigma$ error bars are RSD measurements compiled in Ref.~\cite{kazantzidis2018evolution}. The underlying microphysical parameters are fixed to $\lambda = 10^{-2}$, $v_0 = 1.0\,M_{\mathrm{Pl}}$, and $\xi_{\mathrm{density}} = 10^{-3}$. The dynamical background predicts a systematically suppressed late-time growth amplitude relative to the fixed background, a strong case of the scalar--matter coupling on the background expansion, and the corresponding modification to structure formation. 
}

    \label{fig:predictivesigma}
\end{figure*}

The RSD-only comparison (Fig.~\ref{fig:rsdonly}, Table~\ref{tab:rsd_only_results}) reveals systematic and physically interpretable shifts when the background is treated dynamically. With a fixed \(\Lambda\)CDM background, the marginalized posterior peaks at a relatively low matter fraction, \(\Omega_m \simeq 0.20 \pm 0.09\), and a growth amplitude \(\sigma_{8,0} \simeq 0.75 \pm 0.05\). When the background is solved self-consistently, the posterior moves to a higher matter density \(\Omega_m \simeq 0.31 \pm 0.10\) and a substantially lower growth amplitude \(\sigma_{8,0} \simeq 0.59 \pm 0.01\). The Hubble parameter remains consistent between the two treatments (\(H_0 \simeq 69.8\)--70.2 km\,s\(^{-1}\)Mpc\(^{-1}\) within \(\sim6\) km\,s\(^{-1}\)Mpc\(^{-1}\)), while the coupling parameters \(\beta_0\) and \(a_c\) remain broadly distributed in the fixed-background run and become more tightly localized in the self-consistent run (cf.\ Table~\ref{tab:rsd_only_results}). These shifts indicate that allowing the coupling to modify the background reduces degeneracies between \(\Omega_m\), \(\sigma_{8,0}\) and \(\beta_0\), and that the principal observable impact of the dynamics is a suppression of late-time structure growth.  This is the manifestation at the observable level of the suppressed linear growth induced by the coupling once background feedback is taken into account, as shown in Fig.~\ref{fig:predictivesigma}.

\begin{table}[t]
\scriptsize
\centering
\caption{\small 
RSD-only constraints for the IDE model with an epoch-dependent coupling ($n=1$). The left column corresponds to the fixed $\Lambda$CDM background, while the right column corresponds to the dynamically evolved, self-consistent background in which the scalar field and the Hubble expansion are solved jointly. For the fixed background case, both 68\% and 95\% credible intervals are reported, whereas for the self-consistent background, only the 68\% intervals are shown.
}
\label{tab:rsd_only_results}
\vspace{0.3em}
\begin{tabular}{lccc}
\hline \hline
\textbf{Parameter} & \textbf{Fixed $\Lambda$CDM (68\% C.L.)} & \textbf{Fixed $\Lambda$CDM (95\% C.L.)} & \textbf{Dynamical (68\% C.L.)} \\
\midrule
$\Omega_m$ 
& $0.2035^{+0.0879}_{-0.0867}$ 
& $0.2035^{+0.176}_{-0.172}$ 
& $0.3071^{+0.106}_{-0.097}$ \\[0.3em]

$H_0$ [km\,s$^{-1}$Mpc$^{-1}$] 
& $69.79^{+6.03}_{-5.80}$ 
& $69.79^{+11.9}_{-11.5}$ 
& $70.21^{+6.01}_{-5.81}$ \\[0.3em]

$\sigma_{8,0}$ 
& $0.750^{+0.052}_{-0.051}$ 
& $0.750^{+0.104}_{-0.101}$ 
& $0.590^{+0.011}_{-0.011}$ \\[0.3em]

$\beta_0$ 
& $0.789^{+1.304}_{-0.571}$ 
& $0.789^{+2.582}_{-1.142}$ 
& $0.063^{+0.061}_{-0.056}$ \\[0.3em]

$a_c$ 
& $0.620^{+0.276}_{-0.365}$ 
& $0.620^{+0.548}_{-0.450}$ 
& $0.580^{+0.326}_{-0.323}$ \\
\hline \hline
\end{tabular}
\end{table}

\begin{figure*}
    \centering
    \includegraphics[width=0.7\linewidth]{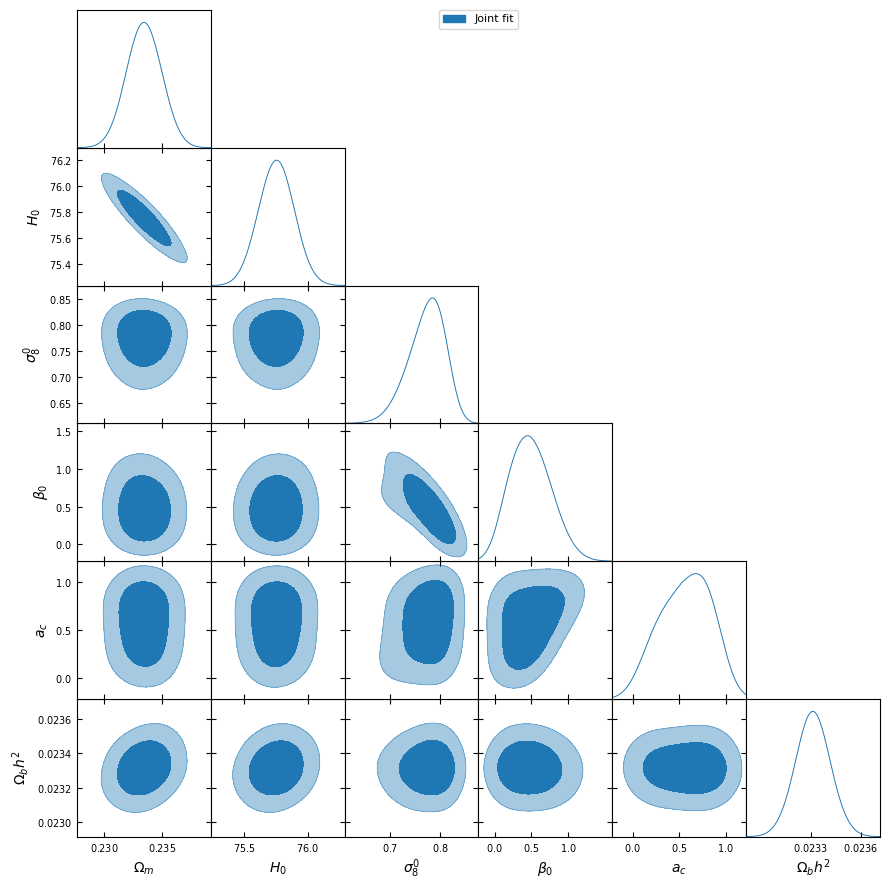}
    \caption{\small Joint posterior distributions for the epoch-dependent coupling model with $n=1$ fixed, using RSD+SN+BAO+CC+\textit{Planck} distance priors. Contours show the 68\% and 95\% credible regions for all two-parameter combinations of $\{\Omega_m, H_0, \sigma_{8,0}, \beta_0, a_c, \Omega_b h^2\}$, with filled regions indicating the marginalized posterior from the joint fit. The standard $\Lambda$CDM parameters $(\Omega_m, H_0, \sigma_{8,0}, \Omega_b h^2)$ are tightly constrained with nearly Gaussian posteriors, while the interaction parameters $(\beta_0, a_c)$ remain weakly constrained: $\beta_0$ is consistent with zero and $a_c$ is broadly distributed. This indicates that current low-redshift data robustly constrain the background sector but have limited sensitivity to the onset and strength of the coupling.}
    \label{fig:jointposter}
\end{figure*}

\begin{figure}
    \centering
    \includegraphics[width=0.7\linewidth]{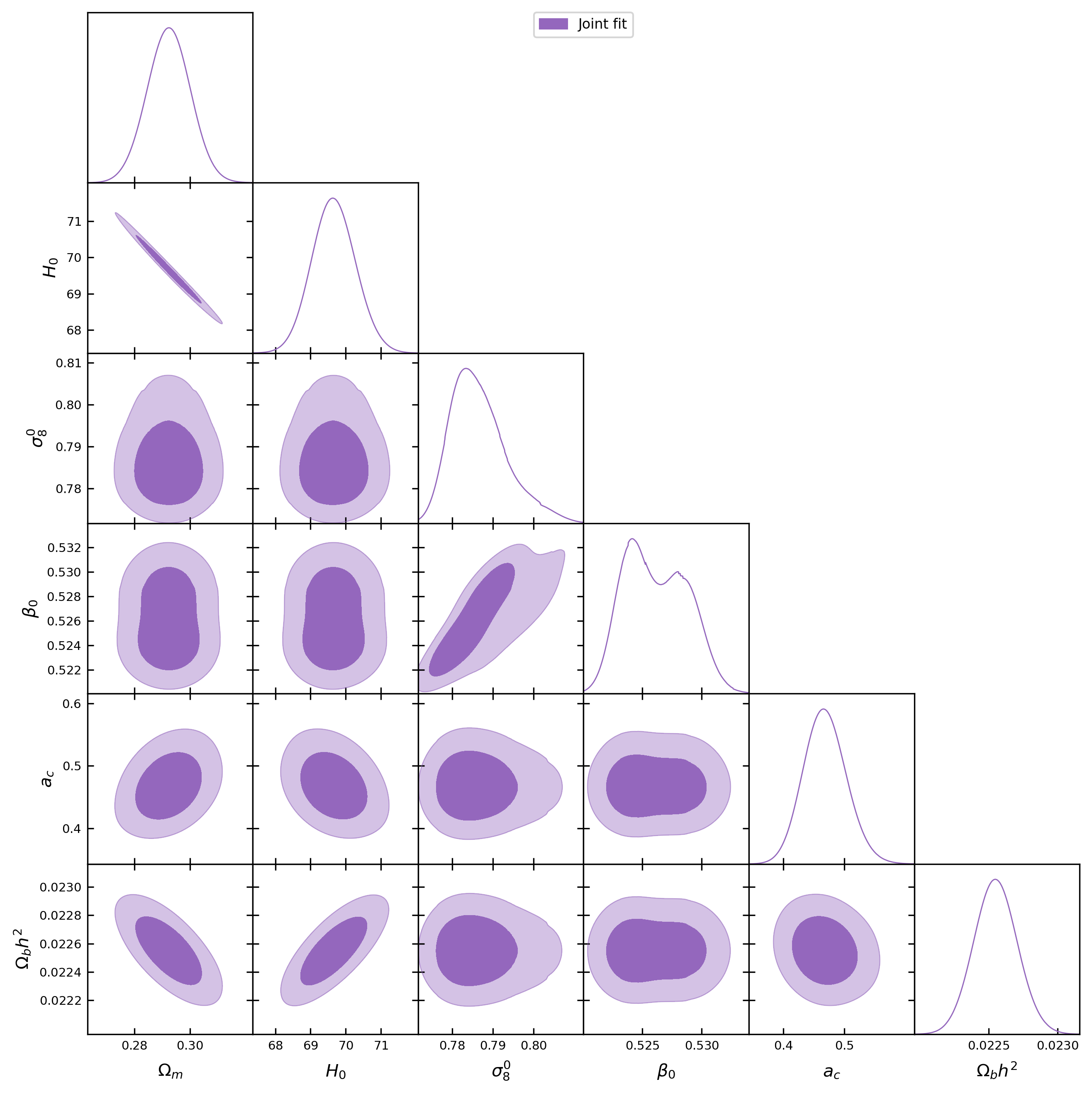}
    \caption{\small Joint posteriors of the dynamical background parameters derived from the full RSD+BAO+CC+Pantheon+SH0ES+\textit{Planck} prior analysis. The Hubble expansion rate $H(a)$ is obtained by integrating the coupled system~\eqref{eq:H_self_consistent} with fixed microphysical parameters $\lambda=0.1$, $v_0=10^{-5}M_{\mathrm{Pl}}$, and $\xi_{\mathrm{density}}=1.0$. The contours (68\% and 95\%) correspond to the dominant posterior components arising from the joint likelihood. For brevity, the subdominant contributions to $\sigma_8$ and $\beta_0$ are shown, highlighting the mild bimodality associated with the late-time activation of the coupling $\beta(a)$.}
    \label{fig:fulliklihood}
\end{figure}

Extending the analysis to the joint likelihood (RSD+BAO+CC+Pantheon+SH0ES with \textit{Planck} 2018 compressed distance priors \((R,\ell_A,\Omega_b h^2)\) following \cite{chen2019distance}) yields substantially tighter and more coherent constraints on the background sector (Figs.~\ref{fig:rsdonly}--\ref{fig:fulliklihood}, Table~\ref{tab:posterior_combined}). In particular, the \textit{Planck}-included chains give \(\Omega_m=0.2334\pm0.0015\), \(H_0 = 75.75\pm0.14\) km\,s\(^{-1}\)Mpc\(^{-1}\) (reflecting the inclusion of the local distance-ladder prior in the joint fit) and \(\sigma_{8,0}=0.77\pm0.04\). The \textit{Planck}-compressed priors strongly reduce parameter degeneracies, but they also render the background tightly constrained in a manner that is prior-driven. In contrast to the RSD-only analysis—where the dynamical-background run favored a nearly vanishing coupling ($\beta_0 \lesssim 0.1$)—the joint likelihood exhibits a statistically significant nonzero coupling, $\beta_0 = 0.524^{+0.0043}_{-0.0008}$, together with a well-localized transition epoch $a_c = 0.467^{+0.035}_{-0.033}$. This indicates that when early- and late-time distance scales are combined with growth data, the model consistently favors a mild but finite dark-sector interaction that falls below the transition redshift $z_c \approx 1.17$. The coupling amplitude is therefore not weakly constrained but rather sharply determined once the background and perturbative observables are jointly analyzed, pointing to a coherent cosmological regime compatible with the \emph{quasi-adiabatic} dynamics discussed earlier.

\begin{figure*}
    \centering
    \includegraphics[width=0.7\linewidth]{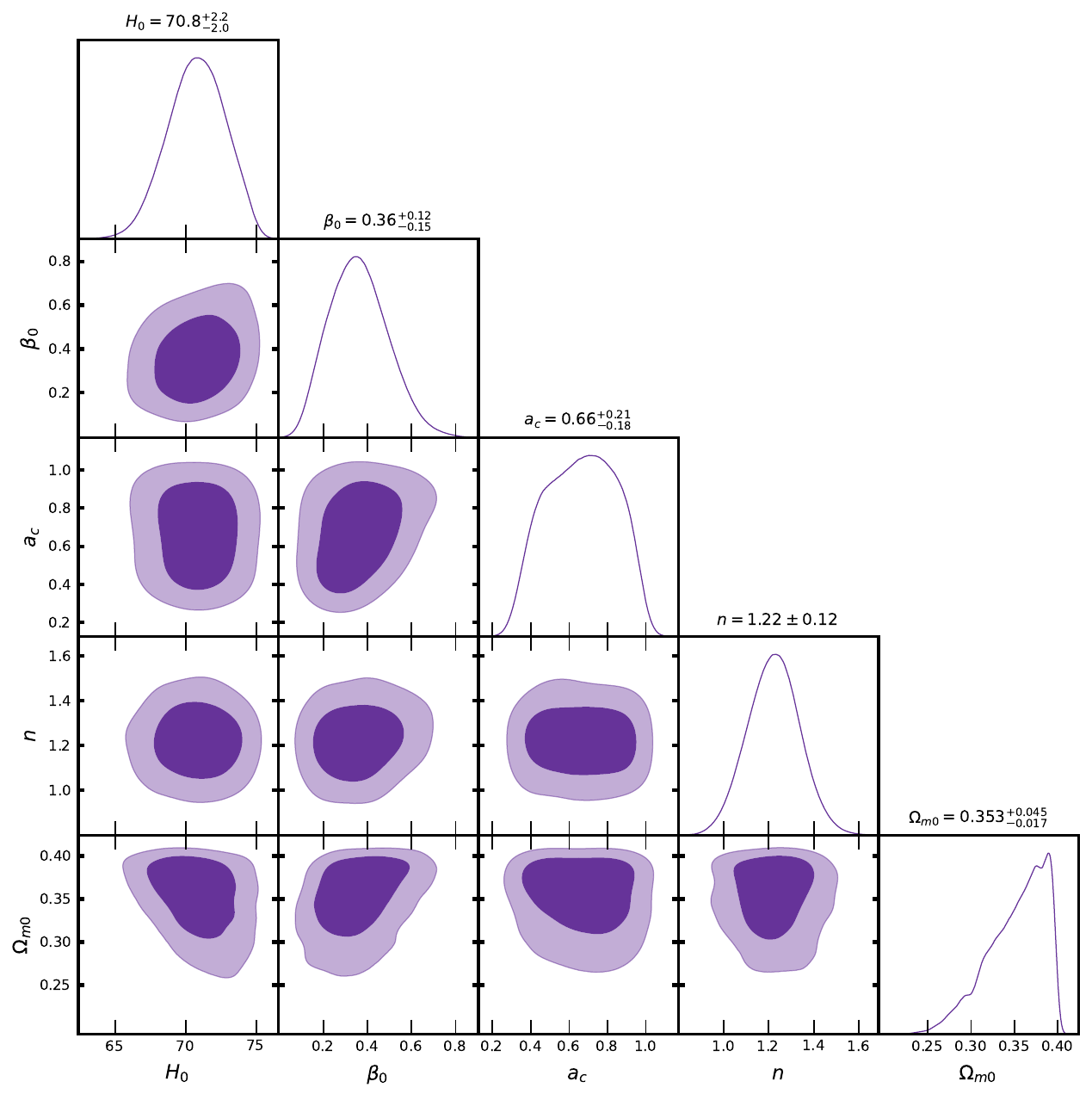}
    \caption{\small 2D likelihood contour plot of the MCMC analysis using CC data with fixed \(\Lambda\)CDM background. The posteriors indicate a mild preference for a nonzero coupling $\beta_0\simeq0.36$ and a smooth transition at $a_c\simeq0.66$ ($z_c\simeq0.5$), consistent with the late-time activation scenario inferred from the RSD analysis.}
    \label{fig:cc}
\end{figure*}

Since our perturbation-level scans freeze the background expansion to the $\Lambda$CDM form, we verified the validity of this approximation (see Supplementary Material II) through an adiabatic diagnostic and exact re-integrations, finding typical interaction strengths \(\Gamma_Q\equiv|Q|/(3H\rho_m)\sim10^{-2}\) and fractional background corrections \(\Delta H/H\lesssim10^{-2}\). The resulting changes in the total likelihood are negligible (\(\Delta\chi^2\lesssim0.5\)), confirming that the frozen-background approximation remains accurate for the \textit{Planck}-constrained ensemble. Independent constraints from CC data provide a complementary, background-level validation of this picture (see Fig. \ref{fig:cc}). The CC analysis yields $H_0 = 70.8^{+2.2}_{-2.0}$ km\,s$^{-1}$Mpc$^{-1}$ and $\Omega_{m0}=0.353^{+0.045}_{-0.017}$, fully consistent with the dynamical-background solution, while mildly favoring a nonzero coupling $\beta_0 = 0.36^{+0.12}_{-0.15}$ and a smooth transition at $a_c \simeq 0.66$ ($z_c \simeq 0.5$). This shows that the coupling remains compatible with direct expansion-rate measurements and that its activation occurs near the onset of cosmic acceleration. However, larger couplings would require a full \textit{Planck}-likelihood run for a definitive early–late consistency test, which is intended for future research.

\begin{table}[htbp]
\scriptsize
\centering
\caption{\small 
Posterior constraints for the IDE model ($n=1$) from the joint RSD+BAO+CC+Pantheon+SH0ES analysis including the compressed \textit{Planck}~2018 distance priors. The left columns correspond to the fixed $\Lambda$CDM background, for which both 68\% and 95\% credible intervals are reported. The right column shows results for the dynamically evolved (self-consistent) background, where the background expansion and the coupling are solved jointly. Quoted values are posterior means with $1\sigma$ (68\%) credible intervals, and 95\% intervals are listed only for the fixed background case.
}
\label{tab:posterior_combined}
\vspace{0.3em}
\begin{tabular}{l c c c}
\hline\hline
\textbf{Parameter} & \textbf{Fixed $\Lambda$CDM (68\% C.L.)} & \textbf{Fixed $\Lambda$CDM  (95\% C.L.)} & \textbf{Dynamical (68\% C.L.)} \\
\hline
$\Omega_m$ 
& $0.2334^{+0.0015}_{-0.0015}$ 
& $0.2334^{+0.0029}_{-0.0029}$ 
& $0.2924^{+0.0075}_{-0.0076}$ \\[0.3em]

$H_0$ [km\,s$^{-1}$Mpc$^{-1}$] 
& $75.76^{+0.13}_{-0.14}$ 
& $75.76^{+0.27}_{-0.27}$ 
& $69.65^{+0.61}_{-0.59}$ \\[0.3em]

$\sigma_{8,0}$ 
& $0.78^{+0.03}_{-0.04}$ 
& $0.78^{+0.05}_{-0.08}$ 
& $0.7844^{+0.0058}_{-0.0045}$ \\[0.3em]

$\beta_0$ 
& $0.47^{+0.30}_{-0.29}$ 
& $0.47^{+0.59}_{-0.44}$ 
& $0.524^{+0.0043}_{-0.0008}$ \\[0.3em]

$a_c$ 
& $0.597^{+0.280}_{-0.348}$ 
& $0.597^{+0.384}_{-0.548}$ 
& $0.467^{+0.035}_{-0.033}$ \\[0.3em]

$\Omega_b h^2$ 
& $0.02330^{+0.00010}_{-0.00010}$ 
& $0.02330^{+0.00020}_{-0.00020}$ 
& $0.02255^{+0.00015}_{-0.00015}$ \\
\hline\hline
\end{tabular}
\end{table}

Physically, there are two notable results: (i) the robust reduction of the growth amplitude when the background dynamics are evolved consistently with the coupling. The \(\sim0.15\) shift in \(\sigma_{8,0}\) between the fixed- and dynamical-background RSD-only runs demonstrates that conclusions about the growth amplitude (and by extension the \(S_8\) tension) are sensitive to whether the coupling is allowed to affect the background. Conversely, \(H_0\) shows minor sensitivity to this modeling choice in the RSD-only analysis.  (ii) When the coupling is small enough that \(\Gamma_Q\ll1\), the dominant observational effect enters through modified perturbation dynamics rather than a significant modification to the expansion history. In this perturbative regime, the coupling enters the growth equation primarily through an effective source and a friction-like term, altering \(f(a)\) and hence \(f\sigma_8(z)\) while leaving the background close to \(\Lambda\)CDM. This is precisely the behavior supported by the \textit{Planck}-included ensemble while leaving background distances largely intact when the coupling is perturbative (see \cite{nesseris2017tension, basilakos2017conjoined}). Conversely, larger couplings would induce non-negligible background backreaction, shifting $H(z)$ and potentially conflicting with \textit{Planck} geometrical constraints.

\section{Discussion and Prospects}

\subsection{Choice of critical epoch}

A defining feature of this model is the existence of a critical epoch $a_c$ at which the effective potential $V_{\mathrm{eff}}(\phi,a)$ transitions from a symmetric to a broken phase, signaling the onset of dark-sector interaction. As the universe expands, the cosmological matter density redshifts as $\rho_c(a)\!\propto\!a^{-3}$, gradually weakening the environmental term in the effective potential. When $\rho_c(a)$ drops below the threshold set by the balance between the potential scale and the density coupling, the curvature at the origin changes sign, and the scalar field condenses into a new VEV. Before this critical epoch, the field remains frozen ($\langle\phi\rangle=0$) and the coupling is inactive, corresponding to a cosmological-constant-like phase. After the transition, the field relaxes toward $\langle\phi\rangle=\pm v_\infty$, inducing an effective Yukawa mass for DM and triggering an energy–momentum exchange through $Q(a)$. The epoch $a_c$ thus marks the moment when cosmic expansion acts as a dynamical analogue of thermal cooling in early-universe physics: the decreasing matter density destabilizes the symmetric vacuum, driving SSB and the activation of the interaction.

From an effective field theory standpoint, the dark sector considered here represents the low-energy limit of an underlying QFT in which the scalar $\phi$ and fermionic $\psi$ interact through a Yukawa coupling. The macroscopic quantities entering the cosmological equations correspond to expectation values of quantum operators in this coarse-grained description, making the scalar effectively an open quantum system weakly coupled to its environment \cite{caldeira1983quantum, boyanovsky2015effective}; represented here by the DM background. Integrating out the fermionic degrees of freedom generates an effective nonlocal influence functional for $\phi$, whose imaginary part induces dissipative and stochastic corrections to its dynamics. At this semiclassical limit, the evolution of $Q(a)$ can be interpreted as a coarse-grained remnant of this quantum backreaction—a dissipative source term arising from the effective interaction between the scalar and its environment. In this interpretation, $a_c$ represents a genuine phase transition between distinct symmetry sectors—analogous to a group contraction from the unbroken $\mathbb{Z}_2$ phase to a broken vacuum manifold \cite{Umezawa1995}. The sharp peak of $\beta(a)$ near $a_c$ and its subsequent decay, observed in the numerical posteriors, support this interpretation as a transient yet cosmologically significant exchange epoch.

\subsection{Future prospects}
The model remains valid in the classical mean-field regime, where the scalar background dominates over quantum fluctuations and higher-order loop corrections are negligible. A complete quantum treatment would require analyzing the renormalizability and radiative stability of the quartic potential, since the density-dependent term introduces environment-dependent running of the coupling parameters. Once symmetry breaking occurs, the resulting Yukawa interaction modifies the fermionic mass, implying that the observed DM mass may be partly generated through scalar condensation rather than primordial freeze-out. This late-time mass acquisition mechanism can imprint mild, redshift-dependent signatures in the growth factor and matter power spectrum, particularly at $z\lesssim1$ \cite{shlivko2024assessing}. These effects are subtle but measurable within the precision of upcoming surveys. Forecasts show that DESI-like redshift-space distortion surveys will significantly constrain time-varying or interacting dark energy couplings \cite{cruickshank2025forecasts}. The wide-area weak-lensing and clustering programs of Euclid and LSST \cite{kumar2025exploring}, in synergy with high-precision CMB missions such as LiteBIRD and CMB-S4 \cite{drewes2024connecting}, promise sub-percent-level growth and expansion constraints at low redshifts that will significantly tighten constraints on the parameter set $\{\beta_0,a_c,n\}$. Together, these observations will determine whether the mild residual growth features identified here represent genuine evidence of a density-driven SSB in the dark sector or can be ruled out as subdominant effects within the standard cosmological model.

\subsection{Conclusion}
Our results show that SSB-induced dark sector interactions can replicate the observed low-redshift growth history without violating background constraints, provided the coupling is activated near the matter–DE equality epoch. Under a fixed \(\Lambda\)CDM background, the posteriors favor small but finite couplings consistent with RSD, BAO, CC, and SN data. When the background is evolved self-consistently, \(H(a)\) deviates by less than one percent at \(z\lesssim1\), converging to the standard expansion at earlier times. This demonstrates that the coupling's dominant observational signatures arise through perturbations—modifying the effective friction and clustering amplitude—rather than through direct changes in the Friedmann geometry. Consequently, the late-time dark-sector interaction can effectively suppress structure growth and relieve the $S_8$ tension, while leaving the background expansion and inferred $H_0$ values essentially unchanged. This separation of influence supports the hypothesis that the $H_0$ and $S_8$ tensions are not manifestations of a single underlying effect but instead trace distinct physical mechanisms acting at different cosmological epochs. The principal observational discriminator will therefore be the interplay between late-time growth observables and the precise CMB acoustic scale. If future data continue to show $\mathcal{O}(10)\sigma$ offsets in the \textit{Planck} distance prior even after including $Q(a)$ self-consistently, the SSB-induced coupling would be strongly disfavored. Conversely, if part of the tension arises from fixing the background to $\Lambda$CDM, a consistent treatment may relax the discrepancy and allow residual couplings at the percent level. Thus, breaking the parameter \((\beta_0, a_c)\) degeneracy in our model requires sub-percent growth measurements and geometric anchors beyond compressed priors. Future work incorporating the full CMB likelihood and non-linear structure formation will determine whether the residual couplings inferred here represent a viable alternative to constant \(\beta\) parameterizations.

Technically, our construction departs from standard symmetron, chameleon, and coupled-quintessence models in a fundamental way. In symmetron and chameleon theories, the scalar field responds to the local ambient density, producing spatially dependent screening that suppresses long-range forces in dense environments. In contrast, the mechanism developed here is global, driven by the mean cosmic DM density and governed by a transition fixed to a cosmological epoch rather than to local environments. Unlike coupled-quintessence models with an effectively constant interaction, the coupling in our framework is an emergent and symmetry-protected quantity that remains negligible during the early universe and becomes dynamically relevant only after symmetry breaking. As a consequence, the background expansion stays close to the concordance model, while modifications arise primarily through perturbations. 

On the theoretical side, extensions such as multifield or Higgs-portal variants that connect dark matter to the same SSB transition, or an effective field theory formulation that organizes higher-order operators and clarifies the ultraviolet behavior, can further generalize the model. Computing fully scale-dependent predictions in regimes where the mediator’s Compton wavelength approaches cosmological scales would also help delineate the limits of the quasi-static approximation and test the robustness of the underlying dark sector dynamics.

\section{Declaration of competing interest}
\begin{enumerate}
    \item Funding: Seed money scheme, Sanction No. CU-ORS-SM-24/29.
    \item Data Availability Statement: all datasets analyzed in this work are publicly available. No new observational data were generated for this study.
    \item Conflicts of Interest: The authors declare no conflict of interest.
\end{enumerate}

\appendix
\section{Linearization of the Interaction Kernel}
\label{app:deltaQ}
In Section \ref{sec:linear_perturbations}, the interaction kernel \(Q\) was introduced as the mediator of energy–momentum exchange between DM and DE. Because this source term enters both the background continuity equations and the first-order perturbation hierarchy, a consistent linearization in conformal time is essential to ensure gauge invariance and correct normalization of the perturbative source.

We therefore begin by defining the conformal-time interaction kernel,

\begin{equation}
\mathcal{Q}(\tau,\mathbf{x}) \equiv a\,Q(t,\mathbf{x}) \;=\; \xi\,\frac{\beta(a)}{M_{\rm Pl}}\,\rho_{\rm DM}(\tau,\mathbf{x})\,\phi'(\tau,\mathbf{x}),
\label{eq:Q_general}
\end{equation}
where $\xi=\pm1$ parametrizes the overall sign convention and primes denote derivatives with respect to conformal time $\tau$. Throughout, we adopt the choice consistent with Sec.~\ref{sec:2}, so that \(Q>0\) (equivalently \(\mathcal{Q}>0\)) corresponds to energy transfer from DM to DE when \(\beta\,\dot{\bar\phi}>0\). The quantity \(\mathcal{Q}\) naturally enters the perturbed hierarchy, while \(Q\) is its cosmic-time analogue in the background system.

\medskip
\noindent\textbf{Linear expansion:} Expanding Eq.~\eqref{eq:Q_general} to first order in perturbations,
\begin{equation}
\phi(\tau,\mathbf{x})=\bar\phi(\tau)+\delta\phi(\tau,\mathbf{x}),
\qquad
\rho_{\rm DM}(\tau,\mathbf{x})
   =\bar\rho_{\rm DM}(\tau)
     \big[1+\delta_{\rm DM}(\tau,\mathbf{x})\big],
\label{eq:decomposed1}
\end{equation}
and defining \(C(\phi)\equiv\beta(\phi)/M_{\rm Pl}\) with 
\(\overline{C}\equiv C(\bar\phi)\) and \(C_{,\phi}\equiv (dC/d\phi)_{\bar\phi}\),
we obtain a linear order
\begin{align}
\mathcal{Q}(\tau,\mathbf{x})
   &= \xi
      \big(\overline{C}+C_{,\phi}\delta\phi\big)
      \bar\rho_{\rm DM}(1+\delta_{\rm DM})
      \big(\bar\phi'+\delta\phi'\big)
      +\mathcal{O}(\delta^2)\nonumber\\
   &= \xi\Big[
      \overline{C}\,\bar\rho_{\rm DM}\,\bar\phi'
      +\overline{C}\,\bar\rho_{\rm DM}\,\delta\phi'
      +\overline{C}\,\bar\rho_{\rm DM}\,\bar\phi'\,\delta_{\rm DM}
      +C_{,\phi}\,\bar\rho_{\rm DM}\,\bar\phi'\,\delta\phi
      \Big]
      +\mathcal{O}(\delta^2).
\end{align}
The first term represents the homogeneous background \(\bar{\mathcal{Q}}\).
The remaining terms constitute the linear perturbation,
\begin{equation}
\delta\mathcal{Q}
   =\xi\Big[
        \overline{C}\,\bar\rho_{\rm DM}\,\delta\phi'
        +\overline{C}\,\bar\rho_{\rm DM}\,\bar\phi'\,\delta_{\rm DM}
        +C_{,\phi}\,\bar\rho_{\rm DM}\,\bar\phi'\,\delta\phi
      \Big].
\label{eq:deltaQ_full}
\end{equation}
The second term on the right-hand side is the standard contribution that appears when \(\beta\) is treated as a background function, whereas the third term (proportional to \(C_{,\phi}\)) represents the additional {mass-shift} correction arising when \(\beta=\beta(\phi)\) and therefore \(\delta\beta\neq0\).

\medskip
\noindent\textbf{Assessing the $C_{,\phi}$ term:} To estimate whether this extra term is relevant, we use the linearized Klein–Gordon equation in Fourier space to relate \(\delta\phi\) to the matter overdensity. The relative contribution of the $C_{,\phi}$ term to the leading terms can be quantified by
\begin{equation}
R(k,a)\equiv\frac{T_{\rm extra}}{T_{\rm lead}}
   \simeq
   \frac{3\,\beta_{,\phi}(a)\,M_{\rm Pl}\,H^2(a)\,\Omega_m(a)}
        {k^2/a^2+m_\phi^2(a)},
\label{eq:supressionfact}
\end{equation}
which serves as a diagnostic ratio. For realistic microphysical realizations, 
\(\beta_{,\phi}\!\sim\!\beta/M_{\rm Pl}\), yielding
\(R(k,a)\simeq 3\beta^2(a)\Omega_m(a)H^2(a)/(k^2/a^2+m_\phi^2(a))\). After symmetry breaking the scalar mass satisfies \(m_\phi\gg k/a\) for all linear cosmological scales, and for benchmark parameters (\(\beta\lesssim1\), \(m_\phi^2/H^2\!\sim\!10^{110}\); see Sec.~\ref{sec:adiabatic}) we obtain \(R\!\ll\!10^{-30}\). Hence, the $C_{,\phi}$ term is completely negligible, justifying the use of the background-evaluated coupling \(\beta(a)\equiv\beta(\phi=v(a))\) in all perturbation equations of the main text.

\medskip
\noindent\textbf{Gauge-consistent linearization:} Adopting the adiabatic (background-only) approximation \(C_{,\phi}\simeq0\) and including the metric perturbation contribution to the DM trace (\(\delta T^{(c)}=\bar\rho_{\rm DM}(\delta_{\rm DM}+\Phi)\)) in Newtonian gauge, we obtain
\begin{equation}
\delta \mathcal{Q}
   = \xi\,\frac{\beta(a)}{M_{\rm Pl}}\,
      \bar\rho_{\rm DM}
      \big[\delta\phi'
           +\bar\phi'(\delta_{\rm DM}+\Phi)\big].
\label{eq:deltaQ_metric_inclusive}
\end{equation}
With the sign choice \(\xi=1\) used in the main text,
\begin{equation}
\delta \mathcal{Q}
   = \frac{\beta(a)}{M_{\rm Pl}}\,
      \bar\rho_{\rm DM}\,
      \big[\bar\phi'(\delta_{\rm DM}+\Phi)+\delta\phi'\big],
\label{eq:deltaQ_final}
\end{equation}
and, if the metric potential \(\Phi\) is neglected for an approximate matter-only treatment,
\begin{equation}
\delta \mathcal{Q}
   = \frac{\beta(a)}{M_{\rm Pl}}\,
      \bar\rho_{\rm DM}\,
      (\bar\phi'\delta_{\rm DM}+\delta\phi').
\label{eq:deltaQ_matteronly}
\end{equation}

Equations~\eqref{eq:deltaQ_full}–\eqref{eq:deltaQ_matteronly} summarize the working forms of the linearized interaction kernel.  
The diagnostic ratio \(R(k,a)\) confirms that neglecting the $C_{,\phi}$ contribution introduces an error far below any observable threshold in the parameter space explored in this work.

\section{Microphysical consistency for SSB-induced dark-sector interactions}
\subsection{ Density--Dependent Effective Mass}
\label{app:densitydep}
In this Appendix, we demonstrate explicitly how the density-dependent effective mass parameter introduced in Eq.~ \eqref{eq:veff_scalar} of the main text arises from a field-theoretic Yukawa coupling between the DE scalar $\phi$ and a fermionic DM field $\psi$. This provides a microscopic underpinning of the expression
\begin{equation}
\mu_{\rm eff}^2(a) = \mu^2 - \xi\,\rho_{\rm DM}(a),
\label{eq:mu_eff_appendix}
\end{equation}
which governs the onset of SSB. Throughout, we work in natural units where $c=\hbar=1$, and coupling constants carry the standard mass dimensions.

At the tree level, we take the scalar potential to have a spontaneously broken $\mathbb{Z}_2$ symmetry,
\begin{equation}
V_0(\phi) = \frac{\lambda}{4}\left(\phi^2 - v_0^2\right)^2,
\label{eq:V0_appendix}
\end{equation}
where $\lambda > 0$ is the self-coupling and $v_0$ is the vacuum expectation value (VEV) in the absence of matter. The curvature at the minimum defines the vacuum scalar mass,
\begin{equation}
\mu^2 \equiv V_0''(v_0) = 2\lambda v_0^2.
\label{eq:mu_tree_appendix}
\end{equation}

We introduce a Yukawa interaction between $\phi$ and $\psi$ of the form
\begin{equation}
\mathcal{L}_{\rm int} = -g\,\phi\,\bar{\psi}\psi,
\label{eq:Lint_appendix}
\end{equation}
where $g$ is the coupling constant. In a homogeneous, non-relativistic  DM background, the fermion bilinear satisfies
\begin{equation}
\langle \bar{\psi}\psi \rangle \simeq \frac{\rho_{\rm DM}}{m_\psi},
\label{eq:bilinear}
\end{equation}
with $m_\psi$ the  DM particle mass and $\rho_{\rm DM}(a) \propto a^{-3}$ the background  DM energy density. This expectation value induces a linear term in the scalar effective potential, giving
\begin{equation}
V_{\rm eff}(\phi;a) = \frac{\lambda}{4}\left(\phi^2 - v_0^2\right)^2 + \frac{g}{m_\psi}\,\rho_{\rm DM}(a)\,\phi.
\label{eq:Veff_appendix}
\end{equation}
The position of the minimum $\phi_{\min}(a)$ is obtained from the stationary condition
\begin{equation}
\lambda\left(\phi_{\min}^2 - v_0^2\right)\phi_{\min} + \frac{g}{m_\psi}\,\rho_{\rm DM}(a) = 0.
\label{eq:stationary_appendix}
\end{equation}

Assuming that the density-induced shift of the VEV is small, $\phi_{\min} = v_0 + \delta$ with $|\delta| \ll v_0$, Eq.~\eqref{eq:stationary_appendix} expands to leading order as
\begin{equation}
\lambda(2v_0\delta)v_0 + \frac{g}{m_\psi}\,\rho_{\rm DM}(a) \simeq 0,
\end{equation}
yielding
\begin{equation}
\delta \simeq -\frac{g\,\rho_{\rm DM}(a)}{2\lambda\,m_\psi\,v_0^2}.
\label{eq:delta_solution_appendix}
\end{equation}
The validity of this expansion requires
\begin{equation}
\frac{g\,\rho_{\rm DM}(a)}{2\lambda\,m_\psi\,v_0^3} \ll 1,
\label{eq:smallness_condition}
\end{equation}
which holds for the parameter ranges considered in the main text, away from the immediate vicinity of the symmetry-breaking transition.

The effective scalar mass-squared is given by the curvature at the shifted minimum,
\begin{equation}
m_\phi^2(a) \equiv V_{\rm eff}''(\phi)\big|_{\phi_{\min}} = \lambda\left(3\phi_{\min}^2 - v_0^2\right).
\label{eq:mphi_def_appendix}
\end{equation}
Using $\phi_{\min} = v_0 + \delta$ and expanding to first order in $\delta$ gives
\begin{equation}
m_\phi^2(a) \simeq 2\lambda v_0^2 + 6\lambda v_0\,\delta.
\end{equation}
Substituting Eq.~\eqref{eq:delta_solution_appendix} into this result yields
\begin{equation}
m_\phi^2(a) \simeq 2\lambda v_0^2 - \frac{3g}{m_\psi\,v_0}\,\rho_{\rm DM}(a) + \mathcal{O}(\rho_{\rm DM}^2).
\label{eq:mphi_rho_appendix}
\end{equation}
Comparing Eqs.~\eqref{eq:mu_tree_appendix} and \eqref{eq:mphi_rho_appendix} gives the density-dependent effective mass in the form of Eq.~\eqref{eq:mu_eff_appendix} with
\begin{equation}
\xi = \frac{3g}{m_\psi\,v_0}.
\label{eq:xi_appendix}
\end{equation}

The sign of $\xi$ is set by the sign of $g$: for $g>0$, increasing $\rho_{\rm DM}$ decreases $m_\phi^2(a)$, and the curvature vanishes at the critical density
\begin{equation}
\rho_\ast = \frac{\mu^2}{\xi} = \frac{2\lambda m_\psi v_0^3}{3g},
\label{eq:rho_star_appendix}
\end{equation}
which defines the symmetry-breaking onset at the scale factor $a_c$ via $\rho_{\rm DM}(a_c) = \rho_\ast$. This mechanism is qualitatively distinct from local screening scenarios such as chameleon models, as it involves a global phase change occurring when the cosmological background density drops below the critical value \(\rho_\ast\). The underlying symmetry is explicitly $\mathbb{Z}_2$, and the effective coupling $\xi$ is determined unambiguously by the microscopic parameters $(g, m_\psi, v_0)$ of the Lagrangian. A close formal analogy is the Coleman-Weinberg mechanism \cite{Morozumi2011Quantum}, in which quantum corrections reshape the potential and induce symmetry breaking. In this case, however, the role of quantum loops is replaced by the classical effect of a finite matter density, leading to a fundamentally different functional dependence of the potential and a distinct physical origin for the transition.

\subsection{Renormalization-scale dependence of the fermion mass}
\label{app:renommass}
The fermion mass entering Eq.~\eqref{eq:beta_yuk} in the main text is the renormalized mass and therefore depends weakly on the choice of renormalization scale \(\mu\). At one loop, the dominant radiative correction from the Yukawa interaction \(\mathcal{L}\supset -y\,\phi\bar\psi\psi\) shifts the fermion pole (or running) mass by a logarithmic term. 
Writing the renormalized mass at scale \(\mu\) as \(m(\mu)=m_{\rm tree}+ \delta m(\mu)\), the one-loop contribution has the schematic form
\begin{equation}
\delta m(\mu)\;=\;\frac{y^2}{16\pi^2}\,m(\mu)\Big[\,A\ln\!\frac{\mu^2}{m(\mu)^2}+B\,\Big] \;+\; \mathcal{O}(y^4),
\label{eq:dm_mu}
\end{equation}
where \(A\) and \(B\) are \(\mathcal{O}(1)\) constants determined by the precise field content, regularization scheme and finite renormalization convention (for a simple real-scalar Yukawa theory one finds \(A\sim -1\) while the anomalous-dimension coefficient entering the running is positive; we keep the general \(A,B\) notation for clarity).  Equivalently, the fermion mass satisfies the renormalization-group equation
\begin{equation}
\mu\frac{d}{d\mu}\ln m(\mu)\;=\;\gamma_m(\mu)
\;\simeq\; \frac{\kappa\,y^2}{16\pi^2} \;+\; \mathcal{O}(y^4),
\label{eq:gamma_m}
\end{equation}
with \(\kappa\) an \(\mathcal{O}(1)\) number (for the minimal Yukawa model \(\kappa=3\) in common conventions).  Integrating Eq.~\eqref{eq:gamma_m} to leading order gives the familiar logarithmic running
\begin{equation}
m(\mu)\simeq m(\mu_0)\left[1+\frac{\kappa\,y^2}{16\pi^2}\ln\!\frac{\mu}{\mu_0}\right] .
\label{eq:m_running}
\end{equation}

Two practical consequences follow for our cosmological parametrization.  First, the combination that appears in Eq.~\eqref{eq:beta_yuk}, \(m_0+y v(a)\), should be interpreted as the renormalized mass evaluated at a scale \(\mu\) chosen to minimize large logarithms (conventionally \(\mu\!\sim\!m_{\rm phys}\simeq m_0+y v\) or \(\mu\!\sim\!v\) after symmetry breaking).  Choosing \(\mu\) close to the physical threshold absorbs the large log in Eq.~\eqref{eq:dm_mu} into the definition of the renormalized mass and keeps \(\delta m\) perturbatively small.  Second, the residual scale dependence provides a simple error estimate: using Eq.~\eqref{eq:m_running}, one finds that the fractional change induced by varying \(\mu\) by an order-unity factor is
\[
\frac{\Delta m}{m}\sim\frac{\kappa\,y^2}{16\pi^2}\ln\!\frac{\mu}{\mu_0},
\]
so that for typical weak Yukawa couplings \(y\lesssim 0.1\) the running is utterly negligible, whereas for \(y\sim\mathcal{O}(1)\) it is at the percent level.

As a concrete illustration, taking $\kappa=3$ and using $\Delta m/m\approx(\kappa y^2/16\pi^2)\ln(\mu/\mu_0)$ for a leading-log estimate, we find
\[
\frac{\kappa y^2}{16\pi^2}\simeq\begin{cases}
1.90\times10^{-4}, & y=0.1,\\[4pt]
1.90\times10^{-2}, & y=1,
\end{cases}
\]
so that a factor-of-two variation in the renormalization scale ($\ln2\simeq0.693$) produces
\[
\frac{\Delta m}{m}\simeq\begin{cases}
1.32\times10^{-4}\ (\simeq0.013\%), & y=0.1,\\[4pt]
1.32\times10^{-2}\ (\simeq1.32\%), & y=1.
\end{cases}
\]
Thus, for perturbative Yukawa couplings $y\lesssim0.1$, the scale dependence is utterly negligible for cosmology; even $y\sim\mathcal{O}(1)$ yields only percent-level shifts. A similar procedure applies to the scalar portal case as well.

For these reasons we adopt the convention \(\mu\approx m_{\rm phys}\) (or \(\mu\approx v\) in the broken phase) when evaluating \(m_0+y v(a)\) in Eq.~\eqref{eq:beta_yuk}; residual scheme- and scale-dependence is then subdominant to the empirical uncertainties on \(\beta_0\) and \((a_c,n)\).  If necessary (for example, if future fits prefer \(y\sim\mathcal{O}(1)\)), a complete two-loop matching and explicit threshold prescription can be implemented to reduce the renormalization-scale uncertainty to the per-mille level.

\section{Declaration of Generative AI and AI-assisted technologies in the writing process}
During the preparation of this work, the author, Pradosh Keshav, utilized Grammarly/ChatGPT-5 to correct grammatical mistakes and enhance the structure of specific paragraphs. After using this tool/service, the author reviewed and edited the content as needed and takes full responsibility for the content of the publication.

\bibliographystyle{unsrt} 
\bibliography{main}

\begin{thebibliography}{100}

\bibitem{riess1998observational}
A.~G. Riess et~al.
\newblock {Observational Evidence from Supernovae for an Accelerating Universe and a Cosmological Constant}.
\newblock {\em Astron. J.}, 116:1009--1038, 1998.

\bibitem{perlmutter1999measurements}
S.~Perlmutter et~al.
\newblock {Measurements of Omega and Lambda from 42 High-Redshift Supernovae}.
\newblock {\em Astrophys. J.}, 517:565--586, 1999.

\bibitem{spergel2003firstyear}
D.~N. Spergel et~al.
\newblock {First-Year Wilkinson Microwave Anisotropy Probe (WMAP) Observations: Determination of Cosmological Parameters}.
\newblock {\em Astrophys. J. Suppl.}, 148:175--194, 2003.

\bibitem{spergel2007threeyear}
D.~N. Spergel et~al.
\newblock {Three-Year Wilkinson Microwave Anisotropy Probe (WMAP) Observations: Implications for Cosmology}.
\newblock {\em Astrophys. J. Suppl.}, 170:377--408, 2007.

\bibitem{tegmark2004cosmological}
M.~Tegmark et~al.
\newblock {Cosmological Parameters from SDSS and WMAP}.
\newblock {\em Phys. Rev. D}, 69:103501, 2004.

\bibitem{abazajian2004first}
K.~Abazajian et~al.
\newblock {The First Data Release of the Sloan Digital Sky Survey}.
\newblock {\em Astron. J.}, 128:502--512, 2004.

\bibitem{abazajian2005third}
K.~Abazajian et~al.
\newblock {The Third Data Release of the Sloan Digital Sky Survey}.
\newblock {\em Astron. J.}, 129:1755--1769, 2005.

\bibitem{riess2022comprehensive}
Adam~G Riess, Wenlong Yuan, Lucas~M Macri, Dan Scolnic, Dillon Brout, Stefano Casertano, David~O Jones, Yukei Murakami, Gagandeep~S Anand, Louise Breuval, et~al.
\newblock A comprehensive measurement of the local value of the hubble constant with 1 km s- 1 mpc- 1 uncertainty from the hubble space telescope and the sh0es team.
\newblock {\em The Astrophysical journal letters}, 934(1):L7, 2022.

\bibitem{Kavya:2025vsj}
N.~S. Kavya, Sai Swagat~Mishra, and P.~K. Sahoo.
\newblock {f(Q) gravity as a possible resolution of the H0 and S8 tensions with DESI DR2}.
\newblock {\em Sci. Rep.}, 15(1):36504, 2025.

\bibitem{Mishra:2025rhi}
Sai~Swagat Mishra, N.~S. Kavya, P.~K. Sahoo, and V.~Venkatesha.
\newblock {Impact of Teleparallelism on Addressing Current Tensions and Exploring the GW Cosmology}.
\newblock {\em Astrophys. J.}, 981(1):13, 2025.

\bibitem{Sudharani:2025cii}
L.~Sudharani, N.~S. Kavya, and V.~Venkatesha.
\newblock {Cosmic structure growth and perturbation analysis in logarithmic $f(Q)$ gravity}.
\newblock {\em Eur. Phys. J. C}, 85(9):997, 2025.

\bibitem{heymans2021kids}
Catherine Heymans, Tilman Tr{\"o}ster, Marika Asgari, Chris Blake, Hendrik Hildebrandt, Benjamin Joachimi, Konrad Kuijken, Chieh-An Lin, Ariel~G S{\'a}nchez, Jan~Luca Van Den~Busch, et~al.
\newblock Kids-1000 cosmology: Multi-probe weak gravitational lensing and spectroscopic galaxy clustering constraints.
\newblock {\em Astronomy \& Astrophysics}, 646:A140, 2021.

\bibitem{mangano2003coupled}
G~Mangano, Gennaro Miele, and V~Pettorino.
\newblock Coupled quintessence and the coincidence problem.
\newblock {\em Modern Physics Letters A}, 18(12):831--842, 2003.

\bibitem{di2017can}
Eleonora Di~Valentino, Alessandro Melchiorri, and Olga Mena.
\newblock Can interacting dark energy solve the h 0 tension?
\newblock {\em Physical Review D}, 96(4):043503, 2017.

\bibitem{di2020interactingde}
Eleonora Di~Valentino, Alessandro Melchiorri, Olga Mena, and Sunny Vagnozzi.
\newblock Interacting dark energy in the early 2020s: A promising solution to the h0 and cosmic shear tensions.
\newblock {\em Physics of the Dark Universe}, 30:100666, 2020.

\bibitem{cai2005cosmic}
R.~G. Cai and A.~Wang.
\newblock {Cosmic dynamics in the Brans-Dicke theory of gravitation}.
\newblock {\em JCAP}, 0503:002, 2005.

\bibitem{olivares2006interacting}
G.~Olivares, F.~Atrio-Barandela, and D.~Pavon.
\newblock {Interacting dark energy and cosmological constraints}.
\newblock {\em Phys. Rev. D}, 74:043521, 2006.

\bibitem{barrow2006interactions}
J.~D. Barrow and T.~Clifton.
\newblock {Cosmologies with energy exchange}.
\newblock {\em Phys. Rev. D}, 73:103520, 2006.

\bibitem{quartin2008}
M.~Quartin, M.~O. Calvao, S.~E. Joras, R.~R.~R. Reis, and I.~Waga.
\newblock {Dark Interactions and Cosmological Fine-Tuning}.
\newblock {\em JCAP}, 0805:007, 2008.

\bibitem{bean2008}
R.~Bean, E.~E. Flanagan, I.~Laszlo, and M.~Trodden.
\newblock {Constraining Interactions in Cosmology's Dark Sector}.
\newblock {\em Phys. Rev. D}, 78:123514, 2008.

\bibitem{valiviita2008large}
Jussi V{\"a}liviita, Elisabetta Majerotto, and Roy Maartens.
\newblock Large-scale instability in interacting dark energy and dark matter fluids.
\newblock {\em Journal of Cosmology and Astroparticle Physics}, 2008(07):020, 2008.

\bibitem{forconi2024double}
Matteo Forconi, William Giar{\`e}, Olga Mena, Eleonora Di~Valentino, Alessandro Melchiorri, Rafael~C Nunes, et~al.
\newblock A double take on early and interacting dark energy from jwst.
\newblock {\em Journal of Cosmology and Astroparticle Physics}, 2024(05):097, 2024.

\bibitem{lin2023dark}
Meng-Xiang Lin, Evan McDonough, J~Colin Hill, and Wayne Hu.
\newblock Dark matter trigger for early dark energy coincidence.
\newblock {\em Physical Review D}, 107(10):103523, 2023.

\bibitem{garcia2024interacting}
Gabriela Garcia-Arroyo, L~Arturo Ure{\~n}a-L{\'o}pez, and J~Alberto V{\'a}zquez.
\newblock Interacting scalar fields: Dark matter and early dark energy.
\newblock {\em Physical Review D}, 110(2):023529, 2024.

\bibitem{Sudharani:2024fdm}
L.~Sudharani, N.~S. Kavya, and V.~Venkatesha.
\newblock {Probing barrow entropy models with future event horizon as IR cutoff}.
\newblock {\em Nucl. Phys. B}, 1009:116725, 2024.

\bibitem{clark2023h}
Steven~J Clark, Kyriakos Vattis, JiJi Fan, and Savvas~M Koushiappas.
\newblock H 0 and s 8 tensions necessitate early and late time changes to $\lambda$ cdm.
\newblock {\em Physical Review D}, 107(8):083527, 2023.

\bibitem{lucca2021dark}
Matteo Lucca.
\newblock Dark energy--dark matter interactions as a solution to the s8 tension.
\newblock {\em Physics of the Dark Universe}, 34:100899, 2021.

\bibitem{naidoo2024dark}
Krishna Naidoo, Mariana Jaber, Wojciech~A Hellwing, and Maciej Bilicki.
\newblock Dark matter solution to the h 0 and s 8 tensions, and the integrated sachs-wolfe void anomaly.
\newblock {\em Physical Review D}, 109(8):083511, 2024.

\bibitem{benisty2024late}
David Benisty, Supriya Pan, Denitsa Staicova, Eleonora Di~Valentino, and Rafael~C Nunes.
\newblock Late-time constraints on interacting dark energy: Analysis independent of h0, rd, and mb.
\newblock {\em Astronomy \& Astrophysics}, 688:A156, 2024.

\bibitem{yashiki2025toward}
Mai Yashiki.
\newblock Toward a simultaneous resolution of the h 0 and s 8 tensions: Early dark energy and an interacting dark sector model.
\newblock {\em Physical Review D}, 112(6):063517, 2025.

\bibitem{Mishra:2025kzu}
Sai~Swagat Mishra and P.~K. Sahoo.
\newblock {Hubble Constant, S8, and Sound Horizon Tensions: A Study Within the Teleparallel Framework}.
\newblock {\em PTEP}, 2025(10):103E03, 2025.

\bibitem{PhysRevD.101.123521}
Abdolali Banihashemi, Nima Khosravi, and Amir~H. Shirazi.
\newblock Phase transition in the dark sector as a proposal to lessen cosmological tensions.
\newblock {\em Phys. Rev. D}, 101:123521, Jun 2020.

\bibitem{garny2024hot}
Mathias Garny, Florian Niedermann, Henrique Rubira, and Martin~S Sloth.
\newblock Hot new early dark energy bridging cosmic gaps: Supercooled phase transition reconciles stepped dark radiation solutions to the hubble tension with bbn.
\newblock {\em Physical Review D}, 110(2):023531, 2024.

\bibitem{di2018vacuum}
Eleonora Di~Valentino, Eric~V Linder, and Alessandro Melchiorri.
\newblock Vacuum phase transition solves the h 0 tension.
\newblock {\em Physical Review D}, 97(4):043528, 2018.

\bibitem{banihashemi2019ginzburg}
Abdolali Banihashemi, Nima Khosravi, and Amir~H Shirazi.
\newblock Ginzburg-landau theory of dark energy: a framework to study both temporal and spatial cosmological tensions simultaneously.
\newblock {\em Physical Review D}, 99(8):083509, 2019.

\bibitem{tan2019dark}
Wanpeng Tan.
\newblock Dark energy and spontaneous mirror symmetry breaking.
\newblock {\em arXiv preprint arXiv:1908.11838}, 2019.

\bibitem{V:2025oex}
Pradosh Keshav~M. V. and Arun Kenath.
\newblock {Quintessence and false vacuum: Two sides of the same coin?}
\newblock {\em Pramana}, 99(2):58, 2025.

\bibitem{khoury2004chameleon}
Justin Khoury and Amanda Weltman.
\newblock Chameleon fields: Awaiting surprises for tests of gravity in space.
\newblock {\em Physical review letters}, 93(17):171104, 2004.

\bibitem{hinterbichler2010screening}
Kurt Hinterbichler and Justin Khoury.
\newblock Screening long-range forces through local symmetry restoration.
\newblock {\em Physical review letters}, 104(23):231301, 2010.

\bibitem{hinterbichler2011symmetron}
Kurt Hinterbichler, Justin Khoury, Aaron Levy, and Andrew Matas.
\newblock Symmetron cosmology.
\newblock {\em Physical Review D—Particles, Fields, Gravitation, and Cosmology}, 84(10):103521, 2011.

\bibitem{Sudharani:2024qnn}
L.~Sudharani, N.~S. Kavya, and V.~Venkatesha.
\newblock {Unveiling the effects of coupling extended Proca-Nuevo gravity on cosmic expansion with recent observations}.
\newblock {\em Mon. Not. Roy. Astron. Soc.}, 535(2):1998--2008, 2024.

\bibitem{pietroni2005dark}
Massimo Pietroni.
\newblock Dark energy condensation.
\newblock {\em Physical Review D—Particles, Fields, Gravitation, and Cosmology}, 72(4):043535, 2005.

\bibitem{olive2008environmental}
Keith~A Olive and Maxim Pospelov.
\newblock Environmental dependence of masses and coupling constants.
\newblock {\em Physical Review D—Particles, Fields, Gravitation, and Cosmology}, 77(4):043524, 2008.

\bibitem{baldi2011clarifying}
Marco Baldi.
\newblock Clarifying the effects of interacting dark energy on linear and non-linear structure formation processes.
\newblock {\em Monthly Notices of the Royal Astronomical Society}, 414(1):116--128, 2011.

\bibitem{baldi2010hydrodynamical}
Marco Baldi, Valeria Pettorino, Georg Robbers, and Volker Springel.
\newblock Hydrodynamical n-body simulations of coupled dark energy cosmologies.
\newblock {\em Monthly Notices of the Royal Astronomical Society}, 403(4):1684--1702, 2010.

\bibitem{Morozumi2011Quantum}
T.~Morozumi, H.~Takata, and Kotaro Tamai.
\newblock Quantum correction to tiny vacuum expectation value in two higgs doublet model for dirac neutrino mass.
\newblock {\em Physical Review D}, page 055002, 2011.

\bibitem{gubitosi2013effective}
Giulia Gubitosi, Federico Piazza, and Filippo Vernizzi.
\newblock The effective field theory of dark energy.
\newblock {\em Journal of Cosmology and Astroparticle Physics}, 2013(02):032, 2013.

\bibitem{bloomfield2013dark}
Jolyon Bloomfield, {\'E}anna~{\'E} Flanagan, Minjoon Park, and Scott Watson.
\newblock Dark energy or modified gravity? an effective field theory approach.
\newblock {\em Journal of Cosmology and Astroparticle Physics}, 2013(08):010, 2013.

\bibitem{gleyzes2014healthy}
J{\'e}r{\^o}me Gleyzes, David Langlois, Federico Piazza, and Filippo Vernizzi.
\newblock Healthy theories beyond horndeski.
\newblock {\em arXiv preprint arXiv:1404.6495}, 2014.

\bibitem{gleyzes2015effective}
J{\'e}r{\^o}me Gleyzes, David Langlois, Michele Mancarella, and Filippo Vernizzi.
\newblock Effective theory of interacting dark energy.
\newblock {\em Journal of Cosmology and Astroparticle Physics}, 2015(08):054, 2015.

\bibitem{gleyzes2016effective}
J{\'e}r{\^o}me Gleyzes, David Langlois, Michele Mancarella, and Filippo Vernizzi.
\newblock Effective theory of dark energy at redshift survey scales.
\newblock {\em Journal of Cosmology and Astroparticle Physics}, 2016(02):056, 2016.

\bibitem{koivisto2005growth}
Tomi Koivisto.
\newblock Growth of perturbations in dark matter coupled with quintessence.
\newblock {\em Physical Review D—Particles, Fields, Gravitation, and Cosmology}, 72(4):043516, 2005.

\bibitem{koivisto2008dynamics}
Tomi Koivisto.
\newblock Dynamics of nonlocal cosmology.
\newblock {\em Physical Review D—Particles, Fields, Gravitation, and Cosmology}, 77(12):123513, 2008.

\bibitem{DAmico2016}
Guido D'Amico, Teresa Hamill, and Nemanja Kaloper.
\newblock Quantum field theory of interacting dark matter and dark energy: Dark monodromies.
\newblock {\em Physical Review D}, 94(10):103526, 2016.

\bibitem{Farrar2004}
Glennys~R. Farrar and P.~James~E. Peebles.
\newblock Interacting dark matter and dark energy.
\newblock {\em The Astrophysical Journal}, 604(1):1--11, 2004.

\bibitem{d2016quantum}
Guido D’Amico, Teresa Hamill, and Nemanja Kaloper.
\newblock Quantum field theory of interacting dark matter and dark energy: Dark monodromies.
\newblock {\em Physical Review D}, 94(10):103526, 2016.

\bibitem{sin1994late}
Sang-Jin Sin.
\newblock Late-time phase transition and the galactic halo as a bose liquid.
\newblock {\em Physical Review D}, 50(6):3650, 1994.

\bibitem{ji1994late}
SU~Ji and Sang-Jin Sin.
\newblock Late-time phase transition and the galactic halo as a bose liquid. ii. the effect of visible matter.
\newblock {\em Physical Review D}, 50(6):3655, 1994.

\bibitem{brax2009decoupling}
Philippe Brax, Carsten van~de Bruck, Anne-Christine Davis, and J{\'e}r{\^o}me Martin.
\newblock Decoupling dark energy from matter.
\newblock {\em Journal of Cosmology and Astroparticle Physics}, 2009(09):032, 2009.

\bibitem{bamba2013spontaneous}
K~Bamba, R~Gannouji, M~Kamijo, S~Nojiri, and M~Sami.
\newblock Spontaneous symmetry breaking in cosmos: The hybrid symmetron as a dark energy switching device.
\newblock {\em Journal of Cosmology and Astroparticle Physics}, 2013(07):017, 2013.

\bibitem{zhang2020obtaining}
Hai-Chao Zhang.
\newblock Obtaining a scalar fifth force via a symmetry-breaking coupling between the scalar field and matter.
\newblock {\em Physical Review D}, 101(4):044020, 2020.

\bibitem{baldi2012multiple}
Marco Baldi.
\newblock Multiple dark matter as a self-regulating mechanism for dark sector interactions.
\newblock {\em Annalen der Physik}, 524(9-10):602--617, 2012.

\bibitem{amendola2004linear}
Luca Amendola.
\newblock Linear and nonlinear perturbations in dark energy models.
\newblock {\em Physical Review D}, 69(10):103524, 2004.

\bibitem{amendola2008quintessence}
Luca Amendola, Marco Baldi, and Christof Wetterich.
\newblock Quintessence cosmologies with a growing matter component.
\newblock {\em Physical Review D—Particles, Fields, Gravitation, and Cosmology}, 78(2):023015, 2008.

\bibitem{bekenstein1993gravitational}
Jacob~D Bekenstein and Robert~H Sanders.
\newblock Gravitational lenses and unconventional gravity theories.
\newblock {\em arXiv preprint astro-ph/9311062}, 1993.

\bibitem{wetterich2003probing}
Christof Wetterich.
\newblock Probing quintessence with time variation of couplings.
\newblock {\em Journal of Cosmology and Astroparticle Physics}, 2003(10):002, 2003.

\bibitem{clemson2012interacting}
Timothy Clemson, Kazuya Koyama, Gong-Bo Zhao, Roy Maartens, and Jussi V{\"a}liviita.
\newblock Interacting dark energy: Constraints and degeneracies.
\newblock {\em Physical Review D—Particles, Fields, Gravitation, and Cosmology}, 85(4):043007, 2012.

\bibitem{pettorino2013testing}
Valeria Pettorino.
\newblock Testing modified gravity with planck: the case of coupled dark energy.
\newblock {\em Physical Review D—Particles, Fields, Gravitation, and Cosmology}, 88(6):063519, 2013.

\bibitem{amendola2000coupled}
Luca Amendola.
\newblock Coupled quintessence.
\newblock {\em Physical Review D}, 62(4):043511, 2000.

\bibitem{amendola2014multifield}
Luca Amendola, Tiago Barreiro, and Nelson~J Nunes.
\newblock Multifield coupled quintessence.
\newblock {\em Physical Review D}, 90(8):083508, 2014.

\bibitem{van2015disformal}
C~van~de Bruck and J~Morrice.
\newblock Disformal couplings and the dark sector of the universe.
\newblock {\em Journal of Cosmology and Astroparticle Physics}, 2015(04):036, 2015.

\bibitem{damour1994string}
Thibault Damour and Alexander~M Polyakov.
\newblock The string dilation and a least coupling principle.
\newblock {\em Nuclear Physics B}, 423(2-3):532--558, 1994.

\bibitem{amendola2000perturbations}
Luca Amendola.
\newblock Perturbations in a coupled scalar field cosmology.
\newblock {\em Monthly Notices of the Royal Astronomical Society}, 312(3):521--530, 2000.

\bibitem{damour1990dark}
Thibault Damour, GW~Gibbons, and C~Gundlach.
\newblock Dark matter, time-varying g, and a dilaton field.
\newblock {\em Physical review letters}, 64(2):123, 1990.

\bibitem{ma1995cosmological}
Chung-Pei Ma and Edmund Bertschinger.
\newblock Cosmological perturbation theory in the synchronous and conformal newtonian gauges.
\newblock {\em arXiv preprint astro-ph/9506072}, 1995.

\bibitem{sawicki2013consistent}
Ignacy Sawicki, Ippocratis~D Saltas, Luca Amendola, and Martin Kunz.
\newblock Consistent perturbations in an imperfect fluid.
\newblock {\em Journal of Cosmology and Astroparticle Physics}, 2013(01):004, 2013.

\bibitem{bean2004probing}
Rachel Bean and Olivier Dore.
\newblock Probing dark energy perturbations: the dark energy equation of state and speed of sound as measured by wmap.
\newblock {\em Physical Review D}, 69(8):083503, 2004.

\bibitem{de2010measuring}
Roland de~Putter, Dragan Huterer, and Eric~V Linder.
\newblock Measuring the speed of dark: Detecting dark energy perturbations.
\newblock {\em Physical Review D—Particles, Fields, Gravitation, and Cosmology}, 81(10):103513, 2010.

\bibitem{pettorino2008coupled}
Valeria Pettorino and Carlo Baccigalupi.
\newblock Coupled and extended quintessence: theoretical differences and structure formation.
\newblock {\em Physical Review D—Particles, Fields, Gravitation, and Cosmology}, 77(10):103003, 2008.

\bibitem{kirzhnits1972macroscopic}
David~A Kirzhnits and Andrei~D Linde.
\newblock Macroscopic consequences of the weinberg model.
\newblock {\em Physics Letters B}, 42(4):471--474, 1972.

\bibitem{kolb1990origin}
Edward~W Kolb, David~S Salopek, and Michael~S Turner.
\newblock Origin of density fluctuations in extended inflation.
\newblock {\em Physical Review D}, 42(12):3925, 1990.

\bibitem{nambu2009nobel}
Yoichiro Nambu.
\newblock Nobel lecture: Spontaneous symmetry breaking in particle physics: A case of cross fertilization.
\newblock {\em Reviews of Modern Physics}, 81(3):1015--1018, 2009.

\bibitem{anderson1963plasmons}
Philip~W Anderson.
\newblock Plasmons, gauge invariance, and mass.
\newblock {\em Physical Review}, 130(1):439, 1963.

\bibitem{englert1964broken}
Fran{\c{c}}ois Englert and Robert Brout.
\newblock Broken symmetry and the mass of gauge vector mesons.
\newblock {\em Physical review letters}, 13(9):321, 1964.

\bibitem{weinberg1967model}
Steven Weinberg.
\newblock A model of leptons.
\newblock {\em Physical review letters}, 19(21):1264, 1967.

\bibitem{brax2011linear}
Philippe Brax, Carsten van~de Bruck, Anne-Christine Davis, Baojiu Li, Benoit Schmauch, and Douglas~J Shaw.
\newblock Linear growth of structure in the symmetron model.
\newblock {\em Physical Review D—Particles, Fields, Gravitation, and Cosmology}, 84(12):123524, 2011.

\bibitem{landim2018dark}
Ricardo~G Landim.
\newblock Dark energy, scalar singlet dark matter and the higgs portal.
\newblock {\em Modern Physics Letters A}, 33(15):1850087, 2018.

\bibitem{altmannshofer2015light}
Wolfgang Altmannshofer, William~A Bardeen, Martin Bauer, Marcela Carena, and Joseph~D Lykken.
\newblock Light dark matter, naturalness, and the radiative origin of the electroweak scale.
\newblock {\em Journal of High Energy Physics}, 2015(1):1--34, 2015.

\bibitem{patt2006higgs}
Brian Patt and Frank Wilczek.
\newblock Higgs-field portal into hidden sectors.
\newblock {\em arXiv preprint hep-ph/0605188}, 2006.

\bibitem{PhysRevLett.118.141802}
Torsten Bringmann, Felix Kahlhoefer, Kai Schmidt-Hoberg, and Parampreet Walia.
\newblock Strong constraints on self-interacting dark matter with light mediators.
\newblock {\em Phys. Rev. Lett.}, 118:141802, Apr 2017.

\bibitem{arcadi2018dark}
Giorgio Arcadi, Miguel~D Campos, Manfred Lindner, Antonio Masiero, and Farinaldo~S Queiroz.
\newblock Dark sequential z' portal: Collider and direct detection experiments.
\newblock {\em Physical Review D}, 97(4):043009, 2018.

\bibitem{arcadi2019real}
Giorgio Arcadi, Oleg Lebedev, Stefan Pokorski, and Takashi Toma.
\newblock Real scalar dark matter: relativistic treatment.
\newblock {\em Journal of High Energy Physics}, 2019(8):1--26, 2019.

\bibitem{coleman1973radiative}
Sidney Coleman and Erick Weinberg.
\newblock Radiative corrections as the origin of spontaneous symmetry breaking.
\newblock {\em Physical Review D}, 7(6):1888, 1973.

\bibitem{pradosh2025loop}
MV~Pradosh~Keshav and Arun Kenath.
\newblock Loop-corrected scalar potentials and late-time acceleration in f (r) gravity.
\newblock {\em The European Physical Journal C}, 85(9):990, 2025.

\bibitem{copeland2006dynamics}
E.~J. Copeland, M.~Sami, and S.~Tsujikawa.
\newblock {Dynamics of Dark Energy}.
\newblock {\em Int. J. Mod. Phys. D}, 15:1753--1936, 2006.

\bibitem{batell2011dark}
Brian Batell.
\newblock Dark discrete gauge symmetries.
\newblock {\em Physical Review D—Particles, Fields, Gravitation, and Cosmology}, 83(3):035006, 2011.

\bibitem{heikinheimo2013twin}
Matti Heikinheimo, Antonio Racioppi, Martti Raidal, and Christian Spethmann.
\newblock Twin peak higgs.
\newblock {\em Physics Letters B}, 726(4-5):781--785, 2013.

\bibitem{camargo2016all}
Jos{\'e}~Eliel Camargo-Molina, Ant{\'o}nio~P Morais, Roman Pasechnik, Marco~OP Sampaio, and Jonas Wess{\'e}n.
\newblock All one-loop scalar vertices in the effective potential approach.
\newblock {\em Journal of High Energy Physics}, 2016(8):1--27, 2016.

\bibitem{upadhye2013symmetron}
Amol Upadhye.
\newblock Symmetron dark energy in laboratory experiments.
\newblock {\em Physical review letters}, 110(3):031301, 2013.

\bibitem{randall2008constraints}
Scott~W Randall, Maxim Markevitch, Douglas Clowe, Anthony~H Gonzalez, and Marusa Brada{\v{c}}.
\newblock Constraints on the self-interaction cross section of dark matter from numerical simulations of the merging galaxy cluster 1e 0657--56.
\newblock {\em The Astrophysical Journal}, 679(2):1173, 2008.

\bibitem{kazantzidis2018evolution}
Lavrentios Kazantzidis and Leandros Perivolaropoulos.
\newblock Evolution of the f $\sigma$ 8 tension with the planck 15/$\lambda$ cdm determination and implications for modified gravity theories.
\newblock {\em Physical Review D}, 97(10):103503, 2018.

\bibitem{jimenez2023cosmic}
Raul Jimenez, Michele Moresco, Licia Verde, and Benjamin~D Wandelt.
\newblock Cosmic chronometers with photometry: a new path to h (z).
\newblock {\em Journal of Cosmology and Astroparticle Physics}, 2023(11):047, 2023.

\bibitem{verde2024tale}
Licia Verde, Nils Sch{\"o}neberg, and H{\'e}ctor Gil-Mar{\'\i}n.
\newblock A tale of many h 0.
\newblock {\em Annual Review of Astronomy and Astrophysics}, 62, 2024.

\bibitem{d2023cosmographic}
Rocco D’Agostino and Rafael~C Nunes.
\newblock Cosmographic view on the h 0 and $\sigma$ 8 tensions.
\newblock {\em Physical Review D}, 108(2):023523, 2023.

\bibitem{moresco20166}
Michele Moresco, Lucia Pozzetti, Andrea Cimatti, Raul Jimenez, Claudia Maraston, Licia Verde, Daniel Thomas, Annalisa Citro, Rita Tojeiro, and David Wilkinson.
\newblock A 6\% measurement of the hubble parameter at z~ 0.45: direct evidence of the epoch of cosmic re-acceleration.
\newblock {\em Journal of Cosmology and Astroparticle Physics}, 2016(05):014, 2016.

\bibitem{Mishra:2025vpy}
Sai~Swagat Mishra, N.~S. Kavya, P.~K. Sahoo, and Tiberiu Harko.
\newblock {Pad{\'e} cosmography and its insights into teleparallel gravity}.
\newblock {\em Mon. Not. Roy. Astron. Soc.}, 543(3):2816--2835, 2025.

\bibitem{delubac2015baryon}
Timoth{\'e}e Delubac, Julian~E Bautista, James Rich, David Kirkby, Stephen Bailey, Andreu Font-Ribera, An{\v{z}}e Slosar, Khee-Gan Lee, Matthew~M Pieri, Jean-Christophe Hamilton, et~al.
\newblock Baryon acoustic oscillations in the ly$\alpha$ forest of boss dr11 quasars.
\newblock {\em Astronomy \& Astrophysics}, 574:A59, 2015.

\bibitem{di2016curvature}
Enea Di~Dio, Francesco Montanari, Alvise Raccanelli, Ruth Durrer, Marc Kamionkowski, and Julien Lesgourgues.
\newblock Curvature constraints from large scale structure.
\newblock {\em Journal of Cosmology and Astroparticle Physics}, 2016(06):013, 2016.

\bibitem{blake2012wigglez}
Chris Blake, Sarah Brough, Matthew Colless, Carlos Contreras, Warrick Couch, Scott Croom, Darren Croton, Tamara~M Davis, Michael~J Drinkwater, Karl Forster, et~al.
\newblock The wigglez dark energy survey: Joint measurements of the expansion and growth history at z< 1.
\newblock {\em Monthly Notices of the Royal Astronomical Society}, 425(1):405--414, 2012.

\bibitem{chuang2013clustering}
Chia-Hsun Chuang, Francisco Prada, Antonio~J Cuesta, Daniel~J Eisenstein, Eyal Kazin, Nikhil Padmanabhan, Ariel~G Sanchez, Xiaoying Xu, Florian Beutler, Marc Manera, et~al.
\newblock The clustering of galaxies in the sdss-iii baryon oscillation spectroscopic survey: single-probe measurements and the strong power of f (z) $\sigma$8 (z) on constraining dark energy.
\newblock {\em Monthly Notices of the Royal Astronomical Society}, 433(4):3559--3571, 2013.

\bibitem{brout2022pantheon+}
Dillon Brout, Georgie Taylor, Dan Scolnic, Charlotte~M Wood, Benjamin~M Rose, Maria Vincenzi, Arianna Dwomoh, Christopher Lidman, Adam Riess, Noor Ali, et~al.
\newblock The pantheon+ analysis: Supercal-fragilistic cross calibration, retrained salt2 light-curve model, and calibration systematic uncertainty.
\newblock {\em The Astrophysical Journal}, 938(2):111, 2022.

\bibitem{brout2022panth+}
Dillon Brout, Dan Scolnic, Brodie Popovic, Adam~G Riess, Anthony Carr, Joe Zuntz, Rick Kessler, Tamara~M Davis, Samuel Hinton, David Jones, et~al.
\newblock The pantheon+ analysis: cosmological constraints.
\newblock {\em The Astrophysical Journal}, 938(2):110, 2022.

\bibitem{scolnic2022pantheon+}
Dan Scolnic, Dillon Brout, Anthony Carr, Adam~G Riess, Tamara~M Davis, Arianna Dwomoh, David~O Jones, Noor Ali, Pranav Charvu, Rebecca Chen, et~al.
\newblock The pantheon+ analysis: the full data set and light-curve release.
\newblock {\em The Astrophysical Journal}, 938(2):113, 2022.

\bibitem{Kavya:2024ssu}
N.~S. Kavya, Sai~Swagat Mishra, P.~K. Sahoo, and V.~Venkatesha.
\newblock {Can teleparallel f(T) models play a bridge between early and late time Universe?}
\newblock {\em Mon. Not. Roy. Astron. Soc.}, 532(3):3126--3133, 2024.

\bibitem{oka2014simultaneous}
Akira Oka, Shun Saito, Takahiro Nishimichi, Atsushi Taruya, and Kazuhiro Yamamoto.
\newblock Simultaneous constraints on the growth of structure and cosmic expansion from the multipole power spectra of the sdss dr7 lrg sample.
\newblock {\em Monthly Notices of the Royal Astronomical Society}, 439(3):2515--2530, 2014.

\bibitem{nesseris2017tension}
Savvas Nesseris, George Pantazis, and Leandros Perivolaropoulos.
\newblock Tension and constraints on modified gravity parametrizations of g eff (z) from growth rate and planck data.
\newblock {\em Physical Review D}, 96(2):023542, 2017.

\bibitem{amendola2004phantom}
Luca Amendola.
\newblock Phantom energy mediates a long-range repulsive force.
\newblock {\em Physical review letters}, 93(18):181102, 2004.

\bibitem{amendola2006constraints}
Luca Amendola, Christos Charmousis, and Stephen~C Davis.
\newblock Constraints on gauss--bonnet gravity in dark energy cosmologies.
\newblock {\em Journal of Cosmology and Astroparticle Physics}, 2006(12):020, 2006.

\bibitem{yoo2012theoretical}
Jaewon Yoo and Yuki Watanabe.
\newblock Theoretical models of dark energy.
\newblock {\em International Journal of Modern Physics D}, 21(12):1230002, 2012.

\bibitem{capozziello2019extended}
Salvatore Capozziello, Rocco D’Agostino, and Orlando Luongo.
\newblock Extended gravity cosmography.
\newblock {\em International Journal of Modern Physics D}, 28(10):1930016, 2019.

\bibitem{raveri2020reconstructing}
Marco Raveri.
\newblock Reconstructing gravity on cosmological scales.
\newblock {\em Physical Review D}, 101(8):083524, 2020.

\bibitem{di2021dark}
Eleonora Di~Valentino, Ankan Mukherjee, and Anjan~A Sen.
\newblock Dark energy with phantom crossing and the h 0 tension.
\newblock {\em entropy}, 23(4):404, 2021.

\bibitem{chen2019distance}
Lu~Chen, Qing-Guo Huang, and Ke~Wang.
\newblock Distance priors from planck final release.
\newblock {\em Journal of Cosmology and Astroparticle Physics}, 2019(02):028, 2019.

\bibitem{basilakos2017conjoined}
Spyros Basilakos and Savvas Nesseris.
\newblock Conjoined constraints on modified gravity from the expansion history and cosmic growth.
\newblock {\em Physical Review D}, 96(6):063517, 2017.

\bibitem{caldeira1983quantum}
Amir~O Caldeira and Anthony~J Leggett.
\newblock Quantum tunnelling in a dissipative system.
\newblock {\em Annals of physics}, 149(2):374--456, 1983.

\bibitem{boyanovsky2015effective}
D~Boyanovsky.
\newblock Effective field theory during inflation: Reduced density matrix and its quantum master equation.
\newblock {\em Physical Review D}, 92(2):023527, 2015.

\bibitem{Umezawa1995}
Hiroomi Umezawa.
\newblock {\em Advanced Field Theory: Micro, Macro, and Thermal Physics}.
\newblock American Institute of Physics, Melville, NY, 1 edition, 1995.
\newblock American Institute of Physics, Melville, NY, published 27 March 1995.

\bibitem{shlivko2024assessing}
David Shlivko and Paul~J Steinhardt.
\newblock Assessing observational constraints on dark energy.
\newblock {\em Physics Letters B}, 855:138826, 2024.

\bibitem{cruickshank2025forecasts}
Nathan Cruickshank, Robert Crittenden, Kazuya Koyama, and Marco Bruni.
\newblock Forecasts for interacting dark energy with time-dependent momentum exchange.
\newblock {\em arXiv preprint arXiv:2504.03555}, 2025.

\bibitem{kumar2025exploring}
Dharmendra Kumar, Ayan Mitra, Shahnawaz~A Adil, and Anjan~A Sen.
\newblock Exploring alternative cosmologies with the lsst: Simulated forecasts and current observational constraints.
\newblock {\em Physical Review D}, 111(4):043503, 2025.

\bibitem{drewes2024connecting}
Marco Drewes and Lei Ming.
\newblock Connecting cosmic inflation to particle physics with litebird, cmb-s4, euclid, and ska.
\newblock {\em Physical Review Letters}, 133(3):031001, 2024.

\end{thebibliography}
\end{document}